\documentclass[sigconf]{acmart}

\usepackage{bm}

\usepackage{xspace}                   
\usepackage{xifthen}
\usepackage{paralist}                 
\usepackage[modulo]{lineno}
\usepackage{soul}                     

\usepackage{siunitx}                 
\expandafter\let\csname ver@natbib.sty\endcsname\relax
\expandafter\let\csname c@author\endcsname\relax
\usepackage[
  backend=bibtex,
  style=numeric,
  maxbibnames=99,
  maxcitenames=3,
  sorting=nyt,
  url=false,
  isbn=false,
  firstinits=true,
  sortcites
]{biblatex}

\renewcommand\citet[1]{\textcite{#1}}
\DeclareLabelalphaTemplate{
  \labelelement{
     \field[final]{shorthand}
     \field{label}
     \field[varwidthnorm,strwidthmax=1]{labelname}
  }
   \labelelement{
     \field[strwidth=2,strside=right]{year}
   }
}

\definecolor{ref-darkblue}{rgb}{0.03,0.3,0.62}
\definecolor{ref-darkorange}{rgb}{1,0.55,0}
\definecolor{ref-turquoise}{rgb}{0.25,0.88,0.82}

\usepackage[capitalize]{cleveref}
\crefname{equation}{Eq.}{Eqs.}
\Crefname{equation}{Equation}{Equations}
\crefname{algocf}{Alg.}{Algs.}
\Crefname{algocf}{Algorithm}{Algorithms}
\crefname{step}{step}{steps}
\Crefname{step}{Step}{Steps}
\crefrangelabelformat{equation}{(#3#1#4--#5#2#6)}
\crefmultiformat{equation}{Eqs.~(#2#1#3)}{ and~(#2#1#3)}{, (#2#1#3)}{, and~(#2#1#3)}
\Crefmultiformat{equation}{Equations~(#2#1#3)}{ and~(#2#1#3)}{, (#2#1#3)}{, and~(#2#1#3)}
\crefname{appendix}{}{} 

\usepackage{relsize}
\def\abbrevsize{.85}
\newcommand\abbrevformat[1]{\textscale{\abbrevsize}{#1}}

\newcommand\abbrev[1]{\abbrevformat{#1}}

\makeatletter
\newcommand{\hypertargetraised}[1]{\Hy@raisedlink{\hypertarget{#1}{}}}
\makeatother

\usepackage{amsthm}

\numberwithin{equation}{section}

\newcommand\para[1]{\paragraph*{#1}}

\usepackage[linesnumbered,lined,ruled,vlined]{algorithm2e}

\usepackage[margin=10pt,font=small,textfont=it,labelfont=bf,justification=justified]{caption}
\newcommand{\algcaption}[3]{
        \ifthenelse{\isempty{#3}}
                   {\caption[#1]{{\sc #2.} \label{#1}}}
                   {\caption[#1]{{\sc #2.} \newline\small{#3} \label{#1}}}
        }

\usepackage{graphicx}
\graphicspath{{figs/}}
\DeclareGraphicsExtensions{.pdf,.png,.gz}
\usepackage[export]{adjustbox}  
\usepackage[
  format=hang,
  singlelinecheck=false,
  font={small},
  labelfont=bf,
  justification=centering,
]{subfig}

\newcommand{\mcaption}[3]{
  \ifthenelse{\isempty{#2}}
             {\caption[#1]{#3 \label{#1}}}
             {\caption[#1]{{\sc #2.} #3 \label{#1}}}
}

\usepackage{tikz,pgfplots,import,tkz-euclide}
\usetikzlibrary{external}
\usetikzlibrary{patterns}

\newif\ifPlotTikz
\PlotTikztrue
\ifPlotTikz
\pgfplotsset{compat=newest}
\usepgfplotslibrary{fillbetween}
\usetikzlibrary{arrows.meta}
\usetikzlibrary{backgrounds}
\usetikzlibrary{pgfplots.groupplots}
\usetikzlibrary{plotmarks}

\pgfplotsset{plot coordinates/math parser=false}
\pgfkeys{/pgf/images/include external/.code=\includegraphics{#1}}
\tikzset{external/export=false}

\newcommand{\includepgf}[2][1]{
\beginpgfgraphicnamed{#2}%
\tikzsetnextfilename{external-#2}%
\scalebox{#1}{\subimport{figs/}{#2.pgf}}%
\endpgfgraphicnamed%
}

\makeatletter
\tikzset{
    every picture/.style={
        execute at begin picture={
            \let\ref\@refstar
        }
    }
}
\makeatother
\else
\newcommand{\includepgf}[2][1]{
\scalebox{#1}{\includegraphics[]{external-#2}}%
}
\fi
\newlength\figureheight
\newlength\figurewidth

\definecolor{plt-blue}{rgb}{0.0078,0.2980,0.7961}
\definecolor{plt-orange}{rgb}{1.0000,0.6431,0.2627}
\definecolor{plt-purple}{rgb}{1.0000,0.2863,0.5255}
\definecolor{plt-violet}{rgb}{0.6118,0.1765,1.0000}

\usepackage{tabularx}     
\usepackage[first=0,last=9]{lcg}
\usepackage{booktabs}     
\usepackage{multirow}
\usepackage{collcell}     
\usepackage{dcolumn}      
\usepackage{ctable}       

\newcommand\ordinal[1]{\ifthenelse{\isin{#1}{abcdefghijklmnopqrstuvwxyz}}{\ensuremath{#1^\mathrm{th}}}{\engordnumber{#1}}}
\newcommand\lbl[1]{\mathrm{#1}}

\let\vector\undefined
\newcommand\mathbfsf[1]{\bm{\mathsf{#1}}}

\newcommand\scalar[1]{#1}

\newcommand\vector[1]{\bm{#1}}
\newcommand\vectord[1]{\mathbfsf{#1}}

\newcommand\reynolds{\ensuremath{R\kern-.06em e}\xspace}  
\newcommand\capillary{\ensuremath{C\kern-.16em a}\xspace} 

\newcommand\vx{\vector{x}}
\newcommand\vB{\vector{B}}
\newcommand\vX{\vector{X}}

\newcommand\vXd{\vectord{X}}
\newcommand\vy{\vector{y}}
\newcommand\vz{\vector{z}}

\newcommand\vn{\vector{n}}
\newcommand\vr{\vector{r}}
\newcommand\vu{\vector{u}}
\newcommand\vg{\vector{g}}

\newcommand\vb{\vector{b}}
\newcommand\vfd{\vectord{f}}
\newcommand\vF{\vectord{F}}

\newcommand\eps{\epsilon}

\newcommand\vf{\vector{f}}

\newcommand\twod{\textscale{.85}{2}\textsc{d}\xspace}
\newcommand\threed{\textscale{.85}{3}\textsc{d}\xspace}

\def\gmres{\abbrev{GMRES}\xspace}
\def\bie{\abbrev{BIE}\xspace}
\def\rbc{\abbrev{RBC}\xspace}

\def\rbcs{\abbrev{RBC}s\xspace}
\def\cpu{\abbrev{CPU}\xspace}

\def\fmm{\abbrev{FMM}\xspace}
\def\pvfmm{\abbrev{PVFMM}\xspace}

\def\lcp{\abbrev{LCP}\xspace}
\def\lcps{\abbrev{LCP}s\xspace}
\def\ncp{\abbrev{NCP}\xspace}

\def\p4est{\texttt{p4est}\xspace}

\def\ib{\abbrev{IB}\xspace}

\def\dpd{\abbrev{DPD}\xspace}
\def\sph{\abbrev{SPH}\xspace}
\def\skx{\abbrev{SKX}\xspace}
\def\knl{\abbrev{KNL}\xspace}
\newcommand\sdc[1][]{\abbrev{SDC\ifthenelse{\isempty{#1}}{}{\kern 1pt}#1}\xspace}
\def\emdash/{\kern 0.2em---\kern 0.2em}

\newcommand\scid[2][1]{\ensuremath{\ifthenelse{\equal{#1}{1}}{}{#1\times}10^{#2}}}

\newcommand\andor[3][-.15em]{#2/{\kern #1}#3}

\usetikzlibrary{fit,backgrounds}

\usepackage{xparse} 
\DeclareDocumentCommand{\nm}{O{n} O{m}}{_{#1#2}}
\DeclareDocumentCommand{\Ynm}{o o}{\ensuremath{\scalar{Y}\nm}}
\def\S2{\ensuremath{{\mathbb S^2}}}

\definecolor{clr1}{RGB}{255, 246, 39}
\definecolor{clr2}{RGB}{124, 22, 28}
\definecolor{clr3}{RGB}{84, 170, 25}
\definecolor{clr4}{RGB}{137, 230, 251}
\definecolor{clr5}{RGB}{226, 49, 39} 
\definecolor{clr6}{RGB}{12, 59, 136}
\definecolor{clr7}{RGB}{53, 120, 120}
\definecolor{clr8}{RGB}{50, 49, 70}
\definecolor{clr9}{RGB}{255, 0, 255}
\definecolor{clr10}{RGB}{0, 0, 255}
\definecolor{clr11}{RGB}{255, 122, 122}

\definecolor{clr12}{RGB}{130, 130, 130}
\definecolor{clr13}{RGB}{180, 180, 180}
\definecolor{clr14}{RGB}{230, 230, 230}

\DeclareDocumentCommand{\Pnm}{O{n} O{m}}{\ensuremath{\scalar{P}_{#1#2}}}

\newcommand{\GridToCoeff}[2]{{{P}}}  
\newcommand{\CoeffToGrid}[2]{{{Q}}}  

\newcommand{\SingularStokes}[1]{{{\bf S}}}   

\begin{document}
\copyrightyear{2019}
\acmYear{2019}
\acmConference[SC '19]{The International Conference for High Performance Computing, Networking, Storage, and Analysis}{November 17--22, 2019}{Denver, CO, USA}
\acmBooktitle{The International Conference for High Performance Computing, Networking, Storage, and Analysis (SC '19), November 17--22, 2019, Denver, CO, USA}
\acmPrice{15.00}
\acmDOI{10.1145/3295500.3356203}
\acmISBN{978-1-4503-6229-0/19/11}

\title[Scalable Red Blood Cell Flows through Vascular Networks]{Scalable Simulation of Realistic Volume Fraction Red Blood Cell Flows through Vascular Networks}

\author{Libin Lu}
\authornote{Both authors contributed equally to this research.}
\affiliation{%
  \institution{Courant Institute of Mathematical Sciences\\ New York University}
  \city{New York}
  \state{NY}
  \postcode{10003}
}
\email{libin@cs.nyu.edu}

\author{Matthew J. Morse}
\authornotemark[1]
\affiliation{%
  \institution{Courant Institute of Mathematical Sciences\\ New York University}
  \city{New York}
  \state{NY}
  \postcode{10003}
}
\email{mmorse@cs.nyu.edu}

\author{Abtin Rahimian}
\affiliation{%
  \institution{Department of Computer Science\\ University of Colorado Boulder}
  \city{Boulder}
  \state{CO}
  \postcode{80302}
}
\email{arahimian@acm.org}

\author{Georg Stadler}
\affiliation{%
  \institution{Courant Institute of Mathematical Sciences\\ New York University}
  \city{New York}
  \state{NY}
  \postcode{10012}
}
\email{stadler@cims.nyu.edu}

\author{Denis Zorin}
\affiliation{%
  \institution{Courant Institute of Mathematical Sciences\\ New York University}
  \city{New York}
  \state{NY}
  \postcode{10003}
}
\email{dzorin@cs.nyu.edu}

\renewcommand{\shortauthors}{Lu and Morse, et al.}

\begin{abstract}
    High-resolution blood flow simulations have potential for developing better understanding biophysical phenomena at the microscale, such as vasodilation, vasoconstriction and overall vascular resistance.
  To this end, we present a scalable platform for the simulation of red blood cell (\rbc) flows through complex capillaries by modeling the physical system as a viscous fluid with immersed deformable particles.
  We describe a parallel boundary integral equation solver for general elliptic partial differential equations, which we apply to Stokes flow through blood vessels.
  We also detail a parallel collision avoiding algorithm to ensure RBCs and the blood vessel remain contact-free.
  We have scaled our code on Stampede2 at the Texas Advanced Computing Center up to 34,816 cores.
  Our largest simulation enforces a contact-free state between four billion
  surface elements and solves for three billion degrees of freedom on 
  one million RBCs and a blood vessel composed from two million patches.
\end{abstract}

\begin{teaserfigure}
\centering
\begin{minipage}[b]{0.48\textwidth}
  \includegraphics[angle=0,width=.98\linewidth]{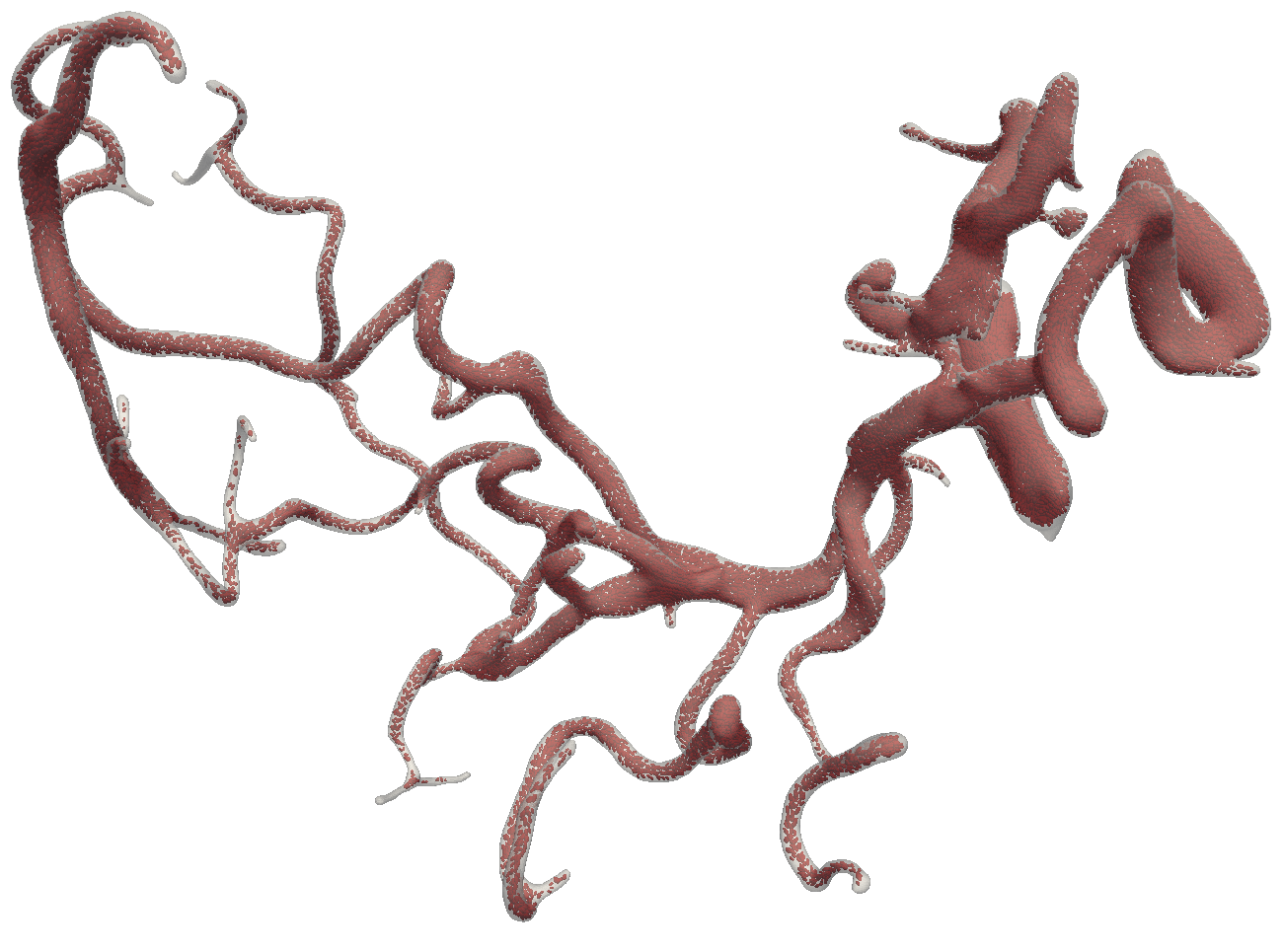}
\end{minipage}
\begin{minipage}[b]{0.48\textwidth}
  \begin{minipage}[b]{0.48\textwidth}
    \includegraphics[angle=0,width=.98\linewidth]{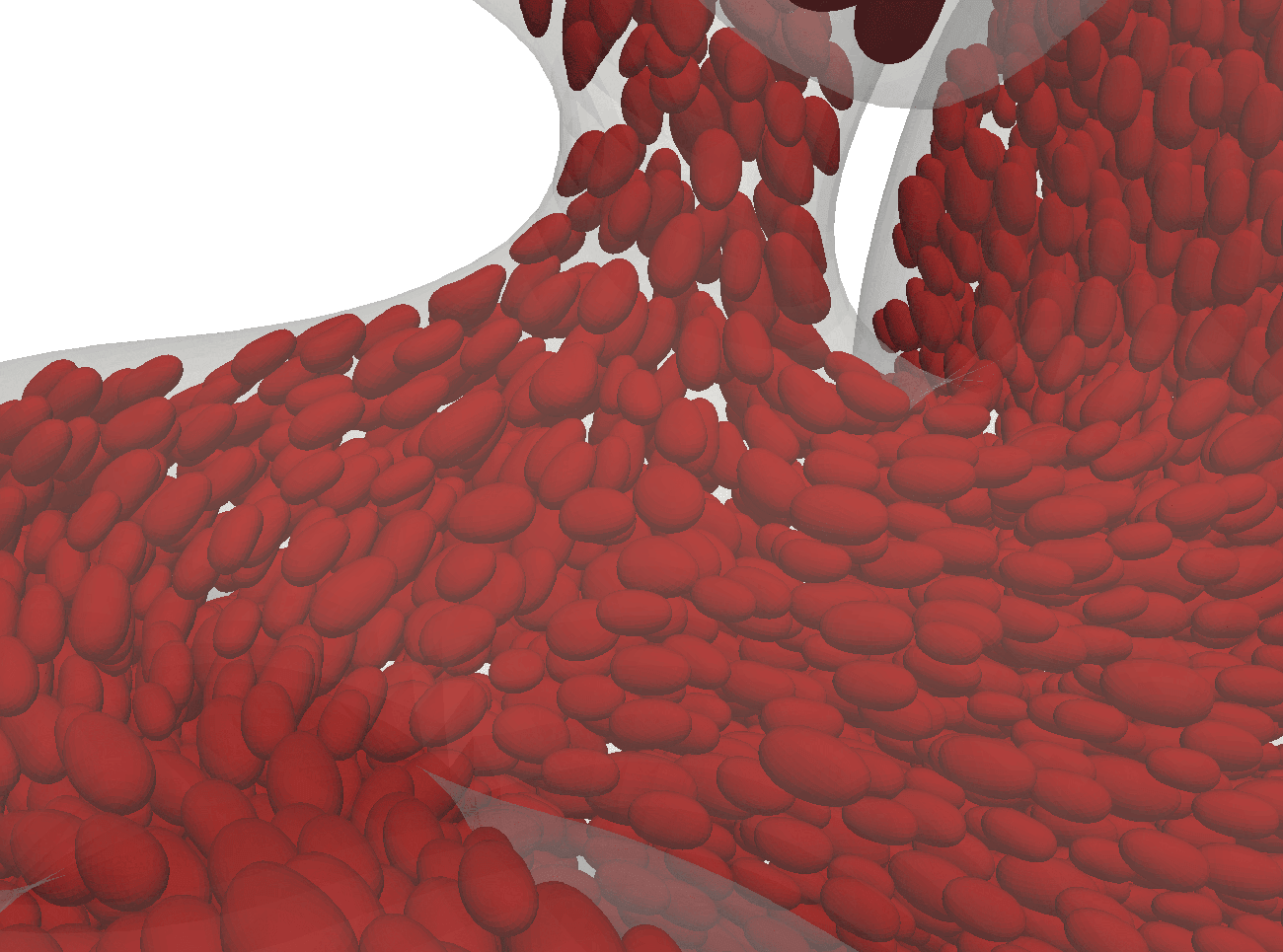}
  \end{minipage}
  \begin{minipage}[b]{0.48\textwidth}
    \includegraphics[angle=0,width=.98\linewidth]{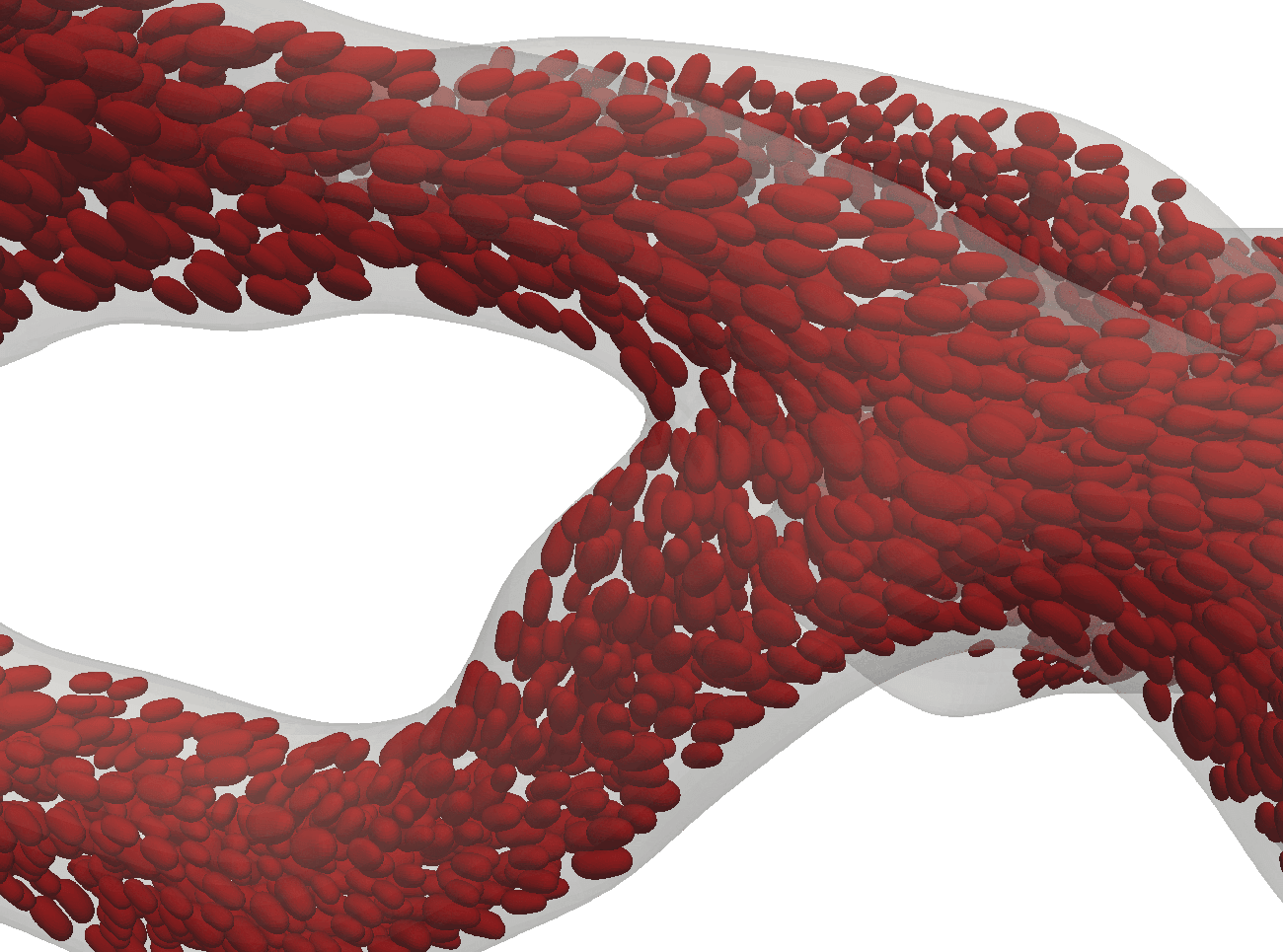}
  \end{minipage}
  \vfill\vspace{.1cm}
  \begin{minipage}[b]{0.48\textwidth}
    \includegraphics[angle=0,width=.98\linewidth]{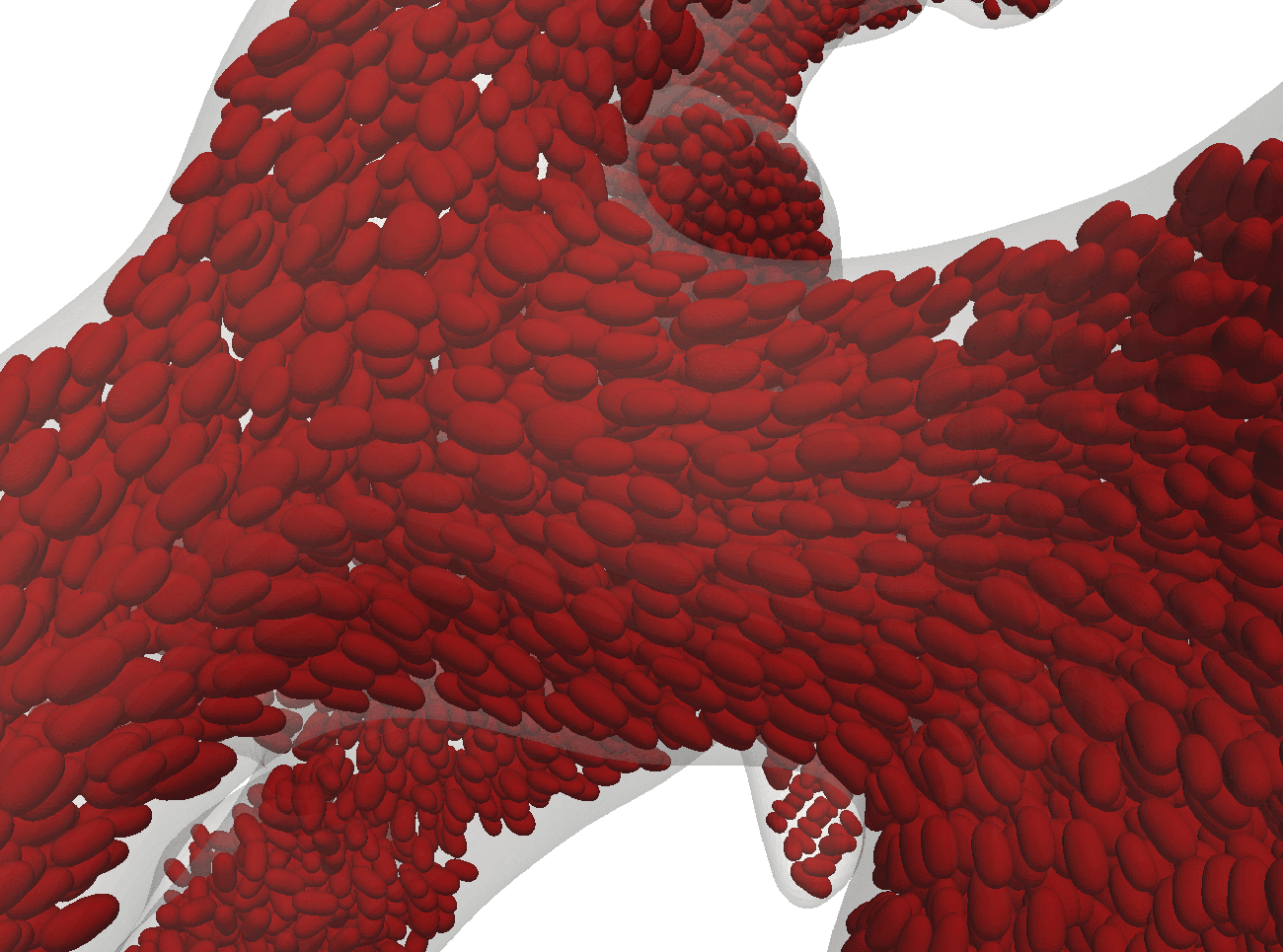}
  \end{minipage}
  \begin{minipage}[b]{0.48\textwidth}
    \includegraphics[angle=0,width=.98\linewidth]{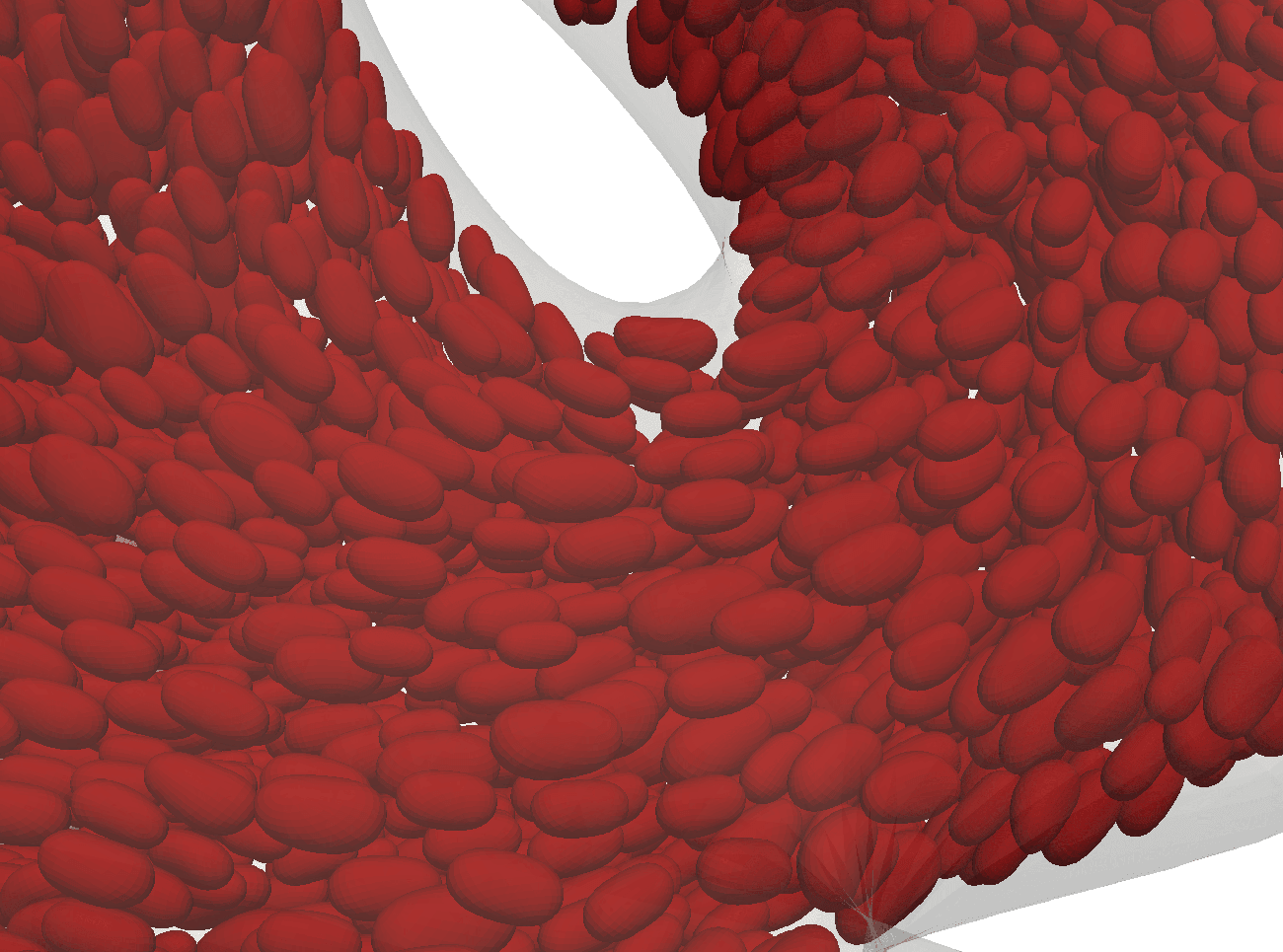}
  \end{minipage}
\end{minipage}
    \mcaption{fig:strong-scale-domain}{}{
    Simulation results for 40,960 \rbcs in a complex vessel geometry.
    For our strong scaling experiments, we use the vessel geometry
    shown on the left, with inflow-outflow boundary conditions at various regions of the vessel geometry.
    To setup the problem, we fill the vessel with 
    nearly-touching \rbcs of different sizes.
    The figure above shows a setup with overall
    40,960 \rbcs at a volume fraction of 19\%, and 40,960 polynomial patches.
    The full simulation video is available at 
    \href{https://vimeo.com/329509229}{\textcolor{blue}{https://vimeo.com/329509229}}.
  }
\end{teaserfigure}


\maketitle

\section{Introduction\label{sec:intro}}
The ability to simulate complex biological flows from first principles
has the potential to provide  insight into complicated physiological processes. 
Simulation of blood flow, in particular, is of paramount biological and clinical importance.
Blood vessel constriction and dilation affects blood pressure, forces between \rbcs can cause clotting, various cells migrate differently through microfluidic devices.

However, direct simulation of blood flow is an extremely challenging task.
Even simulating the blood flow in smaller vessels requires modeling millions of cells (one microliter of blood contains around five million \rbcs) along with a complex blood vessel.
\rbcs are highly deformable and cannot be well-approximated by rigid particles.
The volume fraction of cells in human blood flow reaches 45\%, which means that a very large fraction of cells are in close contact with other cells or vessel walls at any given time.
These constraints preclude a large number of discretization points per cell and make an evolving mesh of the fluid domain impractical and costly at large scale.

Simulations capable of capturing these various types of flows faithfully must be
\begin{itemize}
\item \emph{numerically accurate}, to solve the model equations without
  concern for numerical error;
\item \emph{robust}, to handle high-volume-fraction flows, close contact between cells and vessel walls, complex geometries, and long simulation times;
\item \emph{efficient and scalable}, to support a realistic number of cells in
  flows through complex blood vessels.
\end{itemize}

Achieving these objectives for a blood flow simulation requires that the system meets a number of stringent requirements.
While previous work has made significant progress \cite{Malhotra2017,lu2018parallel,rahimian2010petascale}, we focus on several new infrastructure components essential for handling confined flows and arbitrarily long-time, high volume fractions \rbc flows; in particular, our work is able to realize each of these goals.

We formulate the viscous flow in blood vessels as an integro-differential equation and make use of fast scalable summation algorithms for efficient implementation, as in prior \rbc simulations \cite{Veerapaneni2011}. 
This is the only approach to date that maintains high accuracy at the microscopic level while avoiding expensive discretization of fluid volume: all degrees of freedom reside on the surfaces of \rbcs and blood vessels.


The most important novel aspects of our system include:
(a) handling the \rbc-blood vessel interaction with a fully parallel, high-order
boundary integral equation solver;
(b) explicit handling of collisions with a parallel constraint-based resolution and detection algorithm.
The former is essential for modeling confined flows, while the latter is essential for handling high-volume fraction flows at long time scales without excessively small time steps or fine spatial discretizations. 

\para{Our contributions} 
\begin{enumerate}
  \item We present a parallel platform for long-time simulations of \rbcs through complex blood vessels.
    The extension to suspensions of various particulates (fibers, rigid bodies etc.) is straightforward from the boundary integral formulation. 
    Flows through several complicated geometries are demonstrated.
  \item We have parallelized a boundary solver for elliptic PDEs on smooth complex geometries in \threed. 
    By leveraging the parallel fast-multipole method of \cite{malhotra2015} and the parallel forest of quadtrees of \cite{BursteddeWilcoxGhattas11}, we are able to achieve good parallel performance and load balancing.
  \item We have extended the parallel collision handling of \cite{lu2018parallel} to include rigid \threed boundaries composed of patches. 
  \item We present weak and strong scalability results of our
    simulation on the Skylake cluster and weak scaling results on the
    Knights Landing cluster on Stampede2 at the Texas Advanced
    Computing Center along with several visualizations of long-time,
    large-scale blood cell flows through vessels.
    We observe 
    49\% strong scaling efficiency for a 32-fold increase of compute
    cores. 
    In our largest test on 12288 cores, we simulate 1,048,576 \rbcs in
    a blood vessel composed of 2,097,152
    patches with weak scaling efficiency of 71\% compared to 192 cores
    (\cref{fig:wscale-large-grain}). In each time step, this test uses
    over three billion degrees of freedom
    and over four billion surface elements (triangles) for collision.
  \item We are able to simulate realistic human blood flows with \rbc volume
    fractions over $47\%$ (\cref{fig:high-vol-snap}).

\end{enumerate}
\textbf{Limitations. }
Despite the advantages and contributions of the computational framework presented
here, our work has some limitations.
We have made several simplifications in our model for \rbcs. 
We are restricted to the low Reynolds number regime, i.e., small arteries and capillaries. 
We use a simplified model for \rbcs, assuming the cell membranes to be inextensible
and with no in-plane shear rigidity.
It has been shown that flows in arterioles and capillaries with
diameter of <50 $\mu$m and \rbcs with 5 $\mu$m diameter have a
Reynolds number of $<5 \times 10^{-3}$ \cite{wang2013simulation}\cite[Section 5.4]{caro2012mechanics} with roughly $2$\% error in approximating confined flows \cite{al2008motion}. 
This is sufficient for our interest in the qualitative behavior of particulate flows, with the possibility of investigating rheological dynamics in larger channels.

Regarding algorithms, each \rbc is discretized with an equal number of points, despite the 
varied behavior of the velocity through the vessel. 
Adaptive refinement is required in order to resolve the velocity accurately.
Finally, the blood vessel is constructed to satisfy certain geometric
constraints that allow for the solution of \cref{eq:double_layer_int_eq} via
singular integration.
This can be overcome through uniform refinement, but a parallel adaptive
algorithm is required to maintain good performance. 

\textbf{Related work: blood flow. }
Large-scale simulation of \rbc flows typically fall into four
categories: (a) \textit{Immersed boundary (\ib)} and \textit{immersed
  interface methods}; (b) particle-based methods such as
\textit{dissipative particle dynamics (\dpd)} and \textit{smoothed
  particle hydrodynamics (\sph)} (c) multiscale network-based
approaches and (d) \textit{boundary integral equation (\bie)}
approaches.
For a comprehensive review of general blood flow simulation methods, see \cite{freund2014numerical}.
\ib methods can produce high-quality simulations of heterogeneous particulate flows in complex blood vessels \cite{balogh2017direct,balogh2017computational, xu2013large}.
These methods typically require a finite element solve for each \rbc to compute membrane tensions and use \ib to couple the stresses with the fluid. 
This approach quickly becomes costly, especially for high-order elements, and although reasonably large simulation have been achieved \cite{saadat2018immersed,saadat2019simulation}, large-scale parallelization has remained a challenge.
A different approach to simulating blood flow is with multiscale reduced-order models.
By making simplifying assumptions about the fluid behavior throughout the domain
and transforming the complex fluid system into a simpler flow problem, the
macroscopic behaviors of enormous capillary systems can be characterized 
\cite{peyrounette2018multiscale,perdikaris2016multiscale} and
scaled up to thousands of cores \cite{perdikaris2015effective}.
This comes at a cost of local accuracy; by simulating the flows directly, we are able to accurately resolve local \rbc dynamics that are not captured by such schemes.

Particle-based methods have had the greatest degree of success at large-scale blood flow simulations \cite{gounley2017computational,grinberg2011new,randles2015massively, rossinelli2015silico}. 
These types of approaches are extremely flexible in modeling the fluid and immersed particles, but are computationally demanding and usually suffer from numerical stiffness that requires very small time steps for a given target accuracy. For a comprehensive review, see \cite{ye2016particle}.
There have also been recent advances in coupling a particle-based
\dpd-like scheme with \ib in parallel \cite{ye2017hybrid,ye2018three},
but the number of \rbcs simulated and the complexity of the boundary seems to be limited.

\bie methods have successfully realized large-scale simulations of millions of \rbcs \cite{rahimian2010petascale} in free space.
Recently, new methods for robust handling of collisions between \rbcs in high-volume fraction simulations have been introduced \cite{lu2018parallel,Malhotra2017}. 
This approach is versatile and efficient due to only requiring discretization of
\rbcs and blood vessel surfaces, while achieving high-order convergence and optimal complexity implementation due to fast summation methods \cite{Veerapaneni2009b,Veerapaneni2011,rahimian2015,sorgentone2018highly,sorgentone20193d,af2016fast,Zhao2010}.
To solve elliptic partial differential equations, \bie approaches have been successful in several application domains \cite{YBZ,wala20183d,wala2018optimization,bruno2013high}.
However, to our knowledge, there has been no work combining a Stokes
boundary solver on arbitrary complex geometries in \threed with a
collision detection and resolution scheme to simulate \rbc flows at large scale.
This work aims to fill this gap, illustrating that this can  be
achieved in a scalable manner.

\textbf{Related work: collisions. }
Parallel collision detection methods are a well-studied area in computer graphics for both shared memory and GPU parallelism \cite{Liu2010,Mazhar2011,Kim2009}.
\cite{Iglberger2009,Du2017} detect collisions between rigid bodies in a distributed memory architecture via domain decomposition.
\cite{Pabst2010} constructs a spatial hash to cull collision candidates and explicitly check candidates that hash to the same value.
The parallel geometry and physics-based collision resolution scheme detailed in \cite{yan2019computing} is most similar to the scheme used in this work. 
However, such discrete collision detection schemes require small time steps to guarantee detections which can become costly for high-volume fraction simulations.

\section{Formulation and solver overview\label{sec:formulation}}

\subsection{Problem summary}

We simulate the flow of $N$ cells with deformable boundary surfaces
$\gamma_i$, $i=1,\ldots,N$ in a viscous Newtonian fluid in a domain
$\Omega\subset\mathbb R^3$ with a fixed boundary $\Gamma$. The governing
partial differential equations (PDEs) describing the conservation of
momentum and mass are the incompressible Stokes equations for the
velocity $\vu$ and pressure $p$, combined with velocity boundary
conditions on $\Gamma$.  Additionally, we model cell membranes as massless, so the
velocity $\vXd_t$ of
the points on the cell surface  coincides with the flow velocity:
\begin{align}
  -\mu \Delta \vu(\vx) + \nabla p(\vx) = \vF(\vx) \quad \mathrm{and}\quad \nabla \cdot \vu(\vx) = 0, \quad \vx \in \Omega, 
  \label{eq:stokes_diff1} \\
  \vu(\vx) = \vector g(\vx), \quad \vx \in \Gamma,
  \label{eq:stokes_diff2}  \\
  \vXd_t = \vu(\vXd), \quad \vX \in \gamma_i(t),
  \label{eq:stokes_diff3}  
\end{align}
where $\mu$ is the viscosity of the ambient fluid; in our simulations, we use a simplified model with
the viscosity of the fluid inside the cells also being $\mu$ although
our code supports arbitrary viscosity contrast.  The right-hand side
force in the momentum equation is due to the sum of tension and
bending forces $\vfd = \vfd_\sigma + \vfd_b$; it is concentrated on
the cell surfaces. We assume that cell surfaces are  inextensible,
with bending forces determined by the Canham-Helfrich model \cite{canham1970minimum, helfrich1973elastic},
 based on the surface curvature, and surface tension determined by the surface
incompressibility condition $\nabla_{\gamma_i} \cdot\vu = 0$ resulting in
\[
\vF(\vx) = \sum_i \int_{\gamma_i} \vfd(\vy) \delta(\vx - \vy)d\vy 
\]
(see, e.g., \cite{rahimian2015} for the expressions for $\vfd$).
Except on inflow and outflow regions of the vascular network, the boundary condition $\vector g$ is zero, modeling no-slip boundary condition on blood vessel walls.

\subsubsection{Boundary integral formulation}
To enforce the boundary conditions on $\Gamma$, we use the standard approach of computing
$\vu$ as the sum of the solution $\vu^{\text{fr}}$ of the free-space equation  
\cref{eq:stokes_diff1} without boundary conditions but with non-zero right-hand side $\vF(\vx)$,
and  the second term  $\vu^{\Gamma}$ obtained by solving the
homogeneous equation with boundary conditions on $\Gamma$ given by
$\vector g-\vu^{\text{fr}}$. 

Following the approach of
\cite{Power1987,Poz92,lu2018parallel,nazockdast2015b}, we reformulate
\cref{eq:stokes_diff1,eq:stokes_diff2} in the integral form. The
free-space solution $\vu^{\text{fr}}$ can be written
directly as the sum of the single-layer Stokes potentials $\vu^{\gamma_i}$:

\begin{equation}
  \vu^{\gamma_i}(\vx) = (S_i\vfd)(\vx) = \int_{\gamma_i} S(\vx,\vy)
  \vfd(\vy) d\vy, \quad \vx \in \Omega,
\label{eq:sl_stokes}
\end{equation}
where $S(\vx,\vy) = \frac{1}{8\pi\mu}\left(\frac{1}{\vr} + \frac{\vr \otimes \vr}{|\vr|^3}\right)$ for viscosity $\mu$ and $\vr = \vx - \vy$.

To obtain $\vu^\Gamma$, we reformulate the homogeneous volumetric PDE with nonzero boundary conditions
as a boundary integral equation for an unknown double-layer density $\phi$ defined on the domain boundary $\Gamma$: 
\begin{equation}
  \left(\frac{1}{2}I + D + N \right)\phi = \tilde{D}_\Gamma\phi =  \vector g-\vu^{\text{fr}}, \quad \vx \in \Gamma,
  \label{eq:double_layer_int_eq}\\
\end{equation}
where the double-layer operator is $D\phi(\vx) = \int_\Gamma D(\vx,\vy) \phi(\vy) d\vy$ with double-layer Stokes kernel $D(\vx,\vy) = \frac{6}{8\pi}\left(\frac{\vr \otimes \vr}{|\vr|^5}(\vr\cdot\vn\right)$ for outward normal $\vn=\vn(\vy)$.
The null-space operator needed to make the equations full-rank is defined as
$(N\phi)(\vx) = \int_\Gamma (\vn(\vx) \cdot \phi(\vy))\vn(\vy) d\vy$ (cf.\ \cite{lu2017}).
The favorable eigenspectrum of the integral operator in \cref{eq:double_layer_int_eq} is well-known and allows \gmres to rapidly converge to a solution.
One of the key differences between this work and previous free-space large-scale simulations
is the need to solve this equation in a scalable way. Once the density $\phi$ is computed,
the velocity correction $\vu^\Gamma$ is evaluated directly as $\vu^{\Gamma} = D\phi$.


The  equation for the total velocity $\vu(\vx)$ at any point $\vx \in \Omega$ is then given by
\begin{equation}
  \vu = \vu^{\text{fr}} +\vu^\Gamma =  \sum_{i=1}^N \vu^{\gamma_i} + \vu^\Gamma.
  \label{eq:velocity_combined}
\end{equation}
In particular, this determines the update equation for the boundary
points of cells; see \cref{eq:stokes_diff3}.


\textbf{Contact formulation \label{sec:contact-vol}. }
In theory, the contacts between surfaces are prevented by the increasing fluid forces as surfaces approach each other closely. However, ensuring accuracy of resolving forces may require prohibitively fine sampling of surfaces and very small time steps, making large-scale simulations in space and time impractical. At the same time, as shown in \cite{lu2017}, interpenetration of surfaces results in a catastrophic loss of accuracy due to singularities in the integrals. 

To guarantee that our discretized cells remain interference-free,
we augment \cref{eq:stokes_diff1,eq:stokes_diff2} with an explicit
inequality constraint preventing collisions.  We define a vector
function $V(t)$ with components becoming strictly negative if any cell
surfaces intersect each other, or intersect with the vessel boundaries
$\Gamma$.  More specifically, we use the \emph{space-time interference volumes} introduced in \cite{Harmon2011} and applied to 3D cell flows in \cite{lu2018parallel}.
Each component of $V$ corresponds to a single connected overlap.
The interference-free constraint at time $t$ is then simply $V(t) \geq 0$. 

For this constraint to be satisfied, the forces $\vfd$ are augmented by an artificial collision
force, i.e.,  $\vfd = \vfd_b + \vfd_\sigma + \vfd_c$, $\vfd_c =
\nabla_u V^T \lambda$, where $\lambda$ is the vector of Lagrange
multipliers, which is determined by the additional
\emph{complementarity} conditions:
\begin{equation}
  \lambda(t) \geq 0, \quad V(t) \geq 0, \quad \lambda(t) \cdot V(t) = 0,
  \label{eq:complementarity_prob}
\end{equation}
 at time $t$, where all inequalities are to be understood component-wise.

 To summarize, the system that we solve at every time step can be
 formulated as follows, where we separate equations for different
 cells and global and local parts of the right-hand side, as it is
 important for our time discretization:
\begin{align}
  &\vX_t =  \left(\sum_{j\neq i} S_j \vfd_j  + D \phi\right) +  S_i \vfd_i, \quad \mbox{for points on $\gamma_i$},
  \label{eq:constrained-all1}\\
  &\nabla_{\gamma_i} \cdot \vX_t = 0,\quad    \vfd_j = \vfd(\vX_j,\sigma_j,\lambda),
  \label{eq:constrained-all2} \\
  &B_\Gamma \phi =  \vector g-\sum_{j} S_j \vfd_j, \quad \mbox{for points on $\Gamma$},
  \label{eq:constrained-all3} \\
   &\lambda(t) \geq 0, \quad V(t) \geq 0, \quad \lambda(t) \cdot V(t) = 0.
  \label{eq:constrained-all4}
\end{align}

 At every time step, \eqref{eq:constrained-all4} results in 
coupling of all close $\gamma_i$'s, which requires a non-local computation. 
We follow the approach detailed in \cite{lu2018parallel, lu2017} to define and solve
the \textit{nonlinear complementarity problem} (\ncp) arising from cell-cell
interactions in parallel, and extend it to prevent intersection of cells
with the domain boundary $\Gamma$, as detailed in \cref{sec:parallel-contact}.

\subsection{Algorithm Overview\label{sec:alg_overview}}

Next, we summarize the algorithmic steps used to solve the constrained
integral equations needed to compute cell surface positions and fluid velocities
at each time step.  In the subsequent sections, we detail the
parallel algorithms we developed to obtain good weak and strong scalability, as shown
in \cref{sec:results}.

\textbf{Overall Discretization. } 
\rbc surfaces are discretized using a spherical harmonic
representation, with surfaces sampled uniformly in the standard
latitude-longitude sphere parametrization. The blood vessel surfaces $\Gamma$ are
discretized using a collection of high-order tensor-product polynomial
patches, each sampled at Clenshaw-Curtis quadrature points. The
space-time interference volume function $V(t)$ is computed using a
piecewise-linear approximation as described in \cite{lu2018parallel}.
For time discretization, we use a locally-implicit first order
time-stepping (higher-order time stepping can be easily incorporated).
Interactions between \rbcs and the blood vessel surfaces are computed
\textit{explicitly}, while the self-interaction of a single \rbc is
computed \textit{implicitly}.

The state of the system at every time step is given by a triple of distributed
vectors $(\vX,\sigma, \lambda)$. The first two (cell surface positions and tensions)
are defined at the discretization points of cells. The vector $\lambda$ has variable
length and corresponds to connected components of collision volumes. 
We use the subscript $i$ to denote the subvectors corresponding to $i$-the cell.
$\vX$ and $\sigma$ are solved as a single system that includes the incompressibility
constraint \cref{eq:constrained-all2}.
To simplify exposition, we omit $\sigma$ in our algorithm summary,  which corresponds to
dropping $\vfd_\sigma$  in the Stokes equation, and dropping the surface incompressibility
constraint equation. 

\textbf{Algorithm summary. }
At each step $t$, we compute the new positions $\vX^+_i$ and collision Lagrange multipliers
$\lambda^+$ at time $t^+=t+\Delta t$.  We assume that in the initial configuration there are no collisions,
so the Lagrange multiplier vector $\lambda$ is zero.  Discretizing in
time,
\cref{eq:constrained-all1} becomes
\[ \vX^+_i =  \vX_i + \Delta t\left(\sum_{j\neq i} S_j
\vfd_j(\vX_j,\lambda)  + D \phi(\vX_j, \lambda)\right) +  \Delta t S_i \vfd_i(\vX^+_i, \lambda^+).
\]


At each single time step, we perform the following steps to obtain $(\vX^+, \lambda^+)$ from $(\vX, \lambda)$. Below evaluation of integrals implies using appropriate (smooth, near-singular or singular) quadrature rules on cell or blood vessel surfaces. 

\begin{enumerate}

\item
  Compute the explicit part $\vb$ of the position update (first term in \cref{eq:constrained-all1}).
  \begin{enumerate}
  \item \label{step:rbc_velocity}
    Evaluate $\vu^\lbl{fr}$ from $(\vX, \lambda)$  on $\Gamma$  with
    \cref{eq:sl_stokes}.
  \item\label{step:boundary_solve} Solve \cref{eq:double_layer_int_eq} for the unknown density $\phi$ on $\Gamma$ using GMRES.
  \item\label{step:boundary_evaluate} For each cell, evaluate  $\vu^\Gamma_i = D\phi$ at all cell points $\vX_i$.
  \item\label{step:inter_rbc_evaluate} For each cell $i$, compute the contributions of
    other cells to $\vX_i^+$:  $\vb^c_i = \vu^\lbl{fr} - u^{\gamma_i} = \sum_{j\neq i}S_j\vfd_j$. 
  \item Set $\vb_i = \vu^\Gamma_i + \vb^c_i$.
  \end{enumerate}
  \item \label{step:solve_ncp} Perform the implicit part of the
    update: solve the \ncp obtained  by treating the second
    (self-interaction) term in \cref{eq:constrained-all1} while
    enforcing the complementarity constraints \cref{eq:complementarity_prob}, i.e., solve
    \begin{align}
      \vX^+_i = \vX_i + \Delta t (\vb_i + S_i \vf_i(\vX_i^+,\lambda^+)),\label{eq:ncp1}\\
      \lambda(t^+) \geq 0, \quad V(t^+) \geq 0, \quad \lambda(t^+) \cdot V(t^+) = 0.\label{eq:ncp2}
    \end{align}
\end{enumerate}

\cref{step:rbc_velocity,step:boundary_solve,step:boundary_evaluate,step:inter_rbc_evaluate}
all require evaluation of global integrals, evaluated as sums over quadrature points;
we compute these sums in parallel with \pvfmm. In particular,
\cref{step:boundary_solve} uses \pvfmm as a part of each matrix-vector product in the
GMRES iteration. These matrix-vector product,
as well as \cref{step:rbc_velocity,step:inter_rbc_evaluate,step:boundary_evaluate}
require near-singular integration to compute the velocity accurately
near \rbc and blood vessel surfaces; this requires parallel
communication to find non-local evaluation points.
Details of these computations are discussed
in \cref{sec:solver}.

The \ncp is solved using a sequence of \textit{linear complementarity problems} (\lcp\/s). Algorithmically, this requires parallel searches of collision candidate pairs and the repeated application of the distributed \lcp matrix to distributed
vectors. Details of these computations are provided in \cref{sec:parallel-contact}.

\textbf{Other parallel quadrature methods. }
Various other parallel algorithms are
leveraged to perform boundary integrals for the vessel geometry and \rbcs.
To compute $\vu^{\gamma_i}(\vX)$ for $\vX \in \gamma_i$, the schemes
presented in \cite{Veerapaneni2011} are used to
achieve spectral convergence for single-layer potentials
by performing a spherical harmonic rotation and apply the quadrature rule
of \cite{graham2002fully}.  We use the improved algorithm in
\cite{Malhotra2017} to precompute the singular integration operator
and substantially improve overall complexity.
To compute $\vu^{\gamma_i}(\vX)$ for $\vX$ close to, but not on $\gamma_i$, we follow the
approaches of \cite{sorgentone2018highly, Malhotra2017}, which use a variation of the high-order near-singular
evaluation scheme of \cite{Ying2006}.  Rather than extrapolating the
velocity from nearby check points as in \cref{sec:solver}, we
use \cite{Veerapaneni2011} to
compute the velocity on surface, upsampled quadrature on $\gamma_i$ to compute
the velocity at check points and interpolate the velocity between them to the desired
location.
We mention these schemes for the sake of completeness; they are not the primary contribution of this work, but are critical components of the overall simulation.

\section{Boundary Solver\label{sec:solver}}

The main challenge in incorporating prescribed flow boundary conditions
$\vector g$ on the domain boundary $\Gamma$ is the approximation and
solution of the boundary integral problem
\cref{eq:double_layer_int_eq}. Upon spatial discretization, this is an
extremely large, dense linear system that must be solved at every time
step due to the changing free space solution $\vu^{\text{fr}}$ on the right hand side. Since we aim at a scalable
implementation, we do not assemble the operator on the left hand side
but only implement the corresponding matrix-vector multiplication, i.e., its application to
vectors. Combined with an iterative solver such as \gmres, this matrix-vector multiply is sufficient to solve
\cref{eq:double_layer_int_eq}. Application of the double-layer
operator $D$ to vectors amounts to a near-singular quadrature for
points close to $\Gamma$. Controlling the error in this computation
requires a tailored quadrature scheme. This scheme is detailed below,
where we put a particular emphasis on the challenges due to our 
parallel implementation.

\subsection{Quadrature for integral equation}

The domain boundary $\Gamma$ is given by a collection of
non-overlapping patches $\Gamma = \bigcup_i P_i(Q)$, where $P_i:Q
\rightarrow \mathbb{R}^3$ is defined on $Q=[-1,1]^2$.  We use the
Nystr\"om discretization for \cref{eq:double_layer_int_eq}.  Since $D(\vx,\vy)$
is singular, this requires a singular quadrature scheme for the integral on the
right-hand side.
We proceed in several steps, starting with the direct non-singular
discretization, followed by a distinct discretization for the
singular and near-singular case.

\textbf{Non-singular integral quadrature. }
We discretize the integral in \cref{eq:double_layer_int_eq}, for $\vx
\not\in \Gamma$, by rewriting it as an integral over a set of
patches and then
apply a tensor-product $q$th order Clenshaw-Curtis rule to each patch:
\begin{equation}
 \!\!\vu(\vx) = \sum_i\!\int_{P_i}D(\vx, \vy) \phi(\vy)d\vy_{P_i} \approx \sum_i \!\sum_{j=0}^{q^2} D(\vx,\vy_{ij}) w_{ij}  \phi(\vy_{ij}),
  \label{eq:smooth_double_layer_int_eq_patches_disc}
\end{equation}
where $\vy_{ij} = P_i(\vector{t}_j)$ and $\vector{t}_j\in[-1,1]^2$ is
the $j$th quadrature point and $w_{ij}$ is the
corresponding quadrature weight.
We refer to the points $\vy_{ij}$ as the \textit{coarse discretization
  of\, $\Gamma$} and 
introduce a single global index $\vy_{\ell} = \vy_{ij}$ with $\ell
= \ell(i,j) = (i-1)q^2 + j$, $\ell = 1, \ldots, N$, where $N$ is the total number of quadrature nodes.
We can then rewrite the right-hand side of \eqref{eq:smooth_double_layer_int_eq_patches_disc} compactly as the vector dot product $W(\vx) \cdot \phi$, where  $\phi_\ell = \phi(\vy_\ell)$ and $W_\ell(\vx) = D(\vx,\vy_\ell)w_\ell$ are the quadrature weights in \cref{eq:smooth_double_layer_int_eq_patches_disc}.

As $\vx \to \Gamma$ for $\vx\in \Omega$, the integrand becomes more singular 
and the accuracy of this quadrature rapidly decreases due
to the singularity in the kernel $D$. This requires us to
construct a singular integral discretization for $\vx = \vy_\ell$,
$\ell=1,\ldots ,N$, and general points on $\Gamma$, which is discussed
next. Note that the same method is
used for evaluation of the velocity values at points close to the
surface, once the equation is solved (\emph{near-singular
  integration}).

\textbf{Singular and near-singular integral quadrature. }
\begin{figure}[h]
\centering
\begin{minipage}[b]{0.48\textwidth}
    \centering
    \includegraphics[angle=0,width=.75\linewidth]{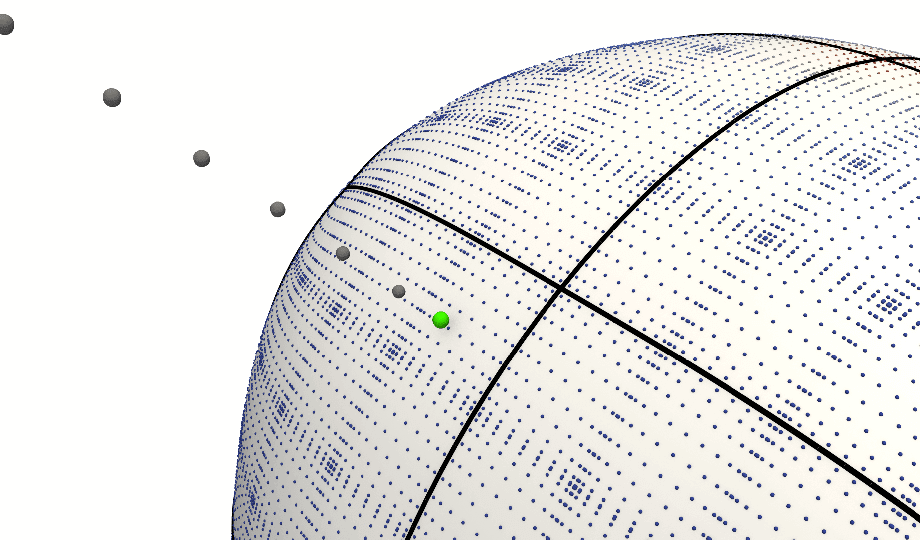}
  \mcaption{fig:hedgehog_schematic}{}{Schematic of our unified
    singular/near-singular quadrature scheme.
    A boundary $\Gamma$ is shown along with a set of patches (patch edges shown in black).
    We evaluate the velocity due to $\Gamma$ at the check points (gray dots off-surface) using the fine discretization (small dots on-surface) and extrapolate these values to the target point (green). 
    The target point may be on or near $\Gamma$.
    The fine discretization subdivides the patches in the coarse discretization into $16$ patches, each with an $11$th-order tensor-product Clenshaw-Curtis quadrature rule.
}
\end{minipage}
\end{figure}
We take an approach similar to \cite{klockner2013quadrature}.
The idea is to evaluate the integral sufficiently far from the surface using the non-singular quadrature rule
\eqref{eq:smooth_double_layer_int_eq_patches_disc} on an upsampled mesh, and then
to extrapolate the accurate values towards the surface.
Concretely, to compute the singular integral at a point $\vx$ near or on
$\Gamma$, we use the following steps:
\begin{enumerate}
  \item Upsample $\phi$ using $q$th order
    interpolation, i.e., $\phi^\lbl{up} = U\phi$, where $\phi^\lbl{up}$ is
    the vector of $Nk$ samples of the density and
    $U$ is the interpolation operator.
    To be precise, we subdivide each patch $P_i$ into $k$ square
    subdomains $P_{ik}$ and use Clenshaw-Curtis nodes in each subdomain.
    We subdivide uniformly, i.e., $P_i$ is split into
    $k=4^\eta$ patches for an integer $\eta$.
    This is the \textit{fine discretization of \,$\Gamma$}. 
    We use $W^\lbl{up}$ to denote the weights for
    \cref{eq:smooth_double_layer_int_eq_patches_disc} the fine discretization quadrature points.
    
  \item Find the closest point $\vy = P(u^*,v^*)$ to $\vx$ on $\Gamma$ for some
  patch $P$ on $\Gamma$ with $u^*,v^*\in [-1,1]$ ($\vy = \vx$ if $\vx \in \Gamma$). 
\item Construct \emph{check points} $c_q =c_q(\vx) = \vy -(R + ir) \vn(u^*,v^*) $,
  $i=0,\ldots, p$, where $\vn(u,v)$ is the outward normal vector to $\Gamma$ at $P(u,v)$. 
  \item Evaluate the velocity at the check points:
    \begin{equation}
      \vu(c_q(\vx)) \approx W^\lbl{up}(c_q)\cdot\phi^{\text{up}},\quad i=0,\ldots, p.
      \label{eq:quad_check_points}
    \end{equation}
  \item Extrapolate the velocity from the check points to $\vx$ with \abbrev{1D} polynomial extrapolation: 
    \begin{align}
      \vu(\vx) \approx& \sum_q e_q \vu(c_q(\vx)) = \left(\sum_q e_q W^\lbl{up}(c_q))\right) U \phi \label{eq:sing-quad}\\
      &= W^\lbl{s}(\vx) \cdot \phi,
    \end{align}      
where $e_q$ are the extrapolation weights.
\end{enumerate}
In this work, the parameters $R,p,r$ and $\eta$ are chosen empirically
to balance the
error in the accuracy of $W^{\text{up}}(c_q) \cdot
\phi^{\text{up}}$ and the extrapolation to
$\vx$. A schematic of this quadrature procedure is shown in
\cref{fig:hedgehog_schematic}.

\textbf{Discretizing the integral equation. }
 With the singular integration method described above, we take $\vx = \vy_\ell$, $\ell = 1\ldots N$, and obtain
the following discretization of \cref{eq:double_layer_int_eq}:
\begin{equation}
  \left(\frac{1}{2}I + A\right) \phi = \vg, \quad A_{\ell m} =
  W_m^\lbl{s}(\vy_\ell) + N_{ij},
  \label{eq:int_eq_disc}
\end{equation}
where $\vg$ is the boundary condition evaluated at
$\vy_\ell$, $W^s_m(\vx)$ is the $m$th component of $W^s(\vx)$ and $N_{ij}$ is the appropriate element of the rank-completing
operator in \cref{eq:double_layer_int_eq}.

The dense operator $A$ is never assembled explicitly.
We use \gmres to solve \cref{eq:int_eq_disc}, which only requires
application of $A$ to vectors $\phi$.
This matrix-vector product is computed using the steps summarized above. 

Extrapolation and upsampling are local computations that are parallelized trivially if all degrees of freedom for each patch are on a single processor.
The main challenges in parallelization of the above singular
evaluation are 1) initially distributing the patches among processors,
2) computing the closest point on $\Gamma$ and 3) evaluating the
velocity at the check points. The parallelization of these
computations is detailed in the remainder of this section.

\textbf{Far evaluation. }
To compute the fluid velocity away from $\Gamma$, where
\cref{eq:double_layer_int_eq} is non-singular, i.e., at the check
points, the integral can be directly evaluated using
\cref{eq:smooth_double_layer_int_eq_patches_disc}. 
Observing that \cref{eq:smooth_double_layer_int_eq_patches_disc} has the form
of an $N$-body summation, we use the \textit{fast-multipole method}
\cite{greengard1987fast} to evaluate it for all target points at once. 
We use the parallel, kernel-independent implementation Parallel Volume Fast Multipole Method (\pvfmm) \cite{malhotra2015,malhotra2016algorithm}, which has been demonstrated to scale to hundreds of thousands of cores. 
\pvfmm handles all of the parallel communication required to aggregate
and distribute the contribution of non-local patches in $O(N)$ time. 

\subsection{Distributing geometry and evaluation parallelization}

We load pieces of the blood vessel geometry, which is provided as a
quad mesh, separately on different processors. Each face of the 
mesh has a corresponding polynomial $P_i$ defining the $i$th patch.

The $k$ levels of patch subdivision induce a uniform quadtree structure within each quad.
We use the \p4est library \cite{BursteddeWilcoxGhattas11} to manage
this surface mesh hierarchy, keep track of neighbor information,
distribute patch data and to refine and coarsen the discretization in
parallel.  
The parallel quadtree algorithms provided by \p4est are used to
distribute the geometry without replicating the complete surface and
polynomial patches across all processors.
\p4est also determines parent-child patch relationships between the coarse and fine discretizations and the coordinates of the child patches to which we interpolate.

Once the geometry is distributed, constructing check points, all necessary information for upsampling and extrapolation are either available on each processor or communicated by \pvfmm. 
This allows these operations to be embarassingly parallel.
  
\subsection{Parallel closest point search}
\label{sec:closest_point}
To evaluate the solution at a point $\vx$, we must find the closest point $\vy$
on the boundary to $\vx$.
The distance $\|\vx -\vy\|_2$ determines whether or not near-singular
integration is required to compute the velocity at $\vx$.
If it is, $\vy$ is used to construct check points. 

In the context of this work, the point $\vx$ is on the surface of an \rbc, which
may be on a different processor than the patch containing $\vy$.
This necessitates a parallel algorithm to search for $\vy$.
%
%
For that purpose, we extend the spatial sorting algorithm from \cite[Algorithm 1]{lu2018parallel} to support our fixed patch-based boundary and detect near pairs of target points and patches.

\begin{enumerate}[a.]
  \item\label{step:bbnear} \textit{Construct a bounding box $B_{P,\eps}$ for the near-zone of each patch.} 
    We choose a distance $d_\eps$ so that for all points $\vz$ further away than $d_\eps$
    from $P$, the quadrature error of integration over $P$ is bounded by $\eps$.
    The set of points closer to $P$ than $d_\eps$ is the \textit{near-zone of
    }$P$.
    We inflate the bounding box $B_P$ of $P$ by $d_\eps$ along the diagonal to
    obtain $B_{P,\eps}$ to contain all such points.
  \item\label{step:computeid} \textit{Sample $B_{P,\eps}$ and compute
    a spatial hash of the samples and $\vx$.}
    Let $H$ be the average diagonal length of all $B_{P,\eps}$.
    We sample the volume contained in $B_{P,\eps}$ with equispaced samples of spacing $h_P < H$.
    Using a spatial hash function, (such as Morton ordering with a spatial grid
    of spacing $H$), we assign hash values to bounding box samples and $\vx$
    to be used as a sorting key.
    This results in a set of hash values that define the near-zone of $\Gamma$.
  \item\label{step:sort} \textit{Sort all samples by the sorting key}. Use the
    parallel sort of \cite{Sundar2013} on the sorting key of bounding box samples and that of $\vx$. 
    This collects all points with identical sorting key (i.e., close positions) and places them on the same processor.    
    If the hash of $\vx$ matches the hash of a bounding box sample,
    then $\vx$ could require near-singular integration, which we check explicitly.
    Otherwise, we can assume $\vx$ is sufficiently far from $P$ and does not require singular integration.
  \item\label{step:distance} \textit{Compute distances $\|\vx - P_i\|$.} For
    each patch $P_i$ with a bounding box key of $\vx$, we locally solve the minimization problem $\min_{(u,v) \in [-1,1]^2} \|\vx - P_i(u,v)\|$ via Newton's method with a backtracking line search.
    This is a local computation since $\vx$ and $P_i$ were communicated during the Morton ID sort.
  \item\label{step:closest} \textit{Choose the closest patch $P_i$}. We perform a global reduce on the distances $\|\vx - P_i\|$ to determine the closest $P_i$ to $\vx$ and communicate back all the relevant information required for singular evaluation back to $\vx$'s processor.
\end{enumerate}


\section{Parallel collision handling}
\label{sec:parallel-contact}
We prevent collisions of \rbcs with other \rbcs and with the vessel
surface $\Gamma$ by solving the \ncp given in \cref{eq:ncp1,eq:ncp2}. 
This is a nonsmooth and non-local problem, whose assembly and efficient solution
is particularly challenging in parallel.
In this section, we summarize our constraint-based approach and algorithm.

We have integrated piecewise polynomial patches into the framework of \cite{lu2018parallel} 
for parallel collision handling, to which we refer the reader for a more detailed discussion. 
The key step to algorithmically unify \rbcs and patches is to \textit{form a linear triangle
mesh approximation} of both objects.
We now want to enforce that these meshes are collision-free subject to the
physics constraints in \cref{eq:ncp1}.

We linearize the \ncp
and solve a sequence of \lcps whose solutions converge
to the \ncp solution.
At a high-level, the collision algorithm proceeds as follows:
\begin{enumerate}
    \item\label{step:near_tri_pair} Find triangle-vertex pairs of distinct meshes that are candidates for collision.
    \item Compute $V(t^+) = V(t^{+,0})$. If any triangle-vertex pairs on distinct meshes collide, the corresponding component of $V(t)$ will be negative.
    \item While $V_i(t^{+,k}) < 0$ for any $i$:
    \begin{enumerate}
        \item Suppose $m$ components of $V(t)$ are negative 
        \item \label{step:lcp_solve} Solve the following linearized version of \cref{eq:ncp1,eq:ncp2}
    \begin{align}
      \vX^{+,k}_i &= \vX_i + \Delta t (\vb_i + S_i(\vfd_i(\vX_i^{+,k},\lambda^{+,k})),\label{eq:ncp1}\\
      \lambda(t^+) \geq 0, &\quad  L(t^{+,k})\geq 0,
      \quad \lambda(t^{+,k}) \cdot L(t^{+,k}) 
      = 0,\label{eq:ncp2}\\
      &\text{where}\quad L(t) = V(t) + \nabla_u V^T \Delta \vX_i(t)
    \end{align}%
    for the $k$th iteration of the loop and $\vX^{+,k}_i = \vX_i +
    \Delta\vX_i(t^{+,k})$. 
  \item Find new candidate triangle-vertex pairs and compute $V(t^{+,k})$.
    \end{enumerate} 
\end{enumerate}
Here, $t^{+,k}$ is the intermediate time step at which a new candidate position $\vX_i^{+,k}$ occurs.
This approach of iteratively solving an \ncp with sequence of \lcps was shown to converge superlinearly in \cite{fang1984linearization}. 
In \cite{yan2019computing}, the authors demonstrate that one \lcp linearization can approximate the \ncp accurately; our algorithm uses around seven \lcp solves to approximately solve the \ncp.
Upon convergence of this algorithm, we are guaranteed that our system is collision-free.

\begin{figure}[h]
\centering
\begin{minipage}[b]{0.48\textwidth}
    \centering
    \includegraphics[angle=0,width=.7\linewidth]{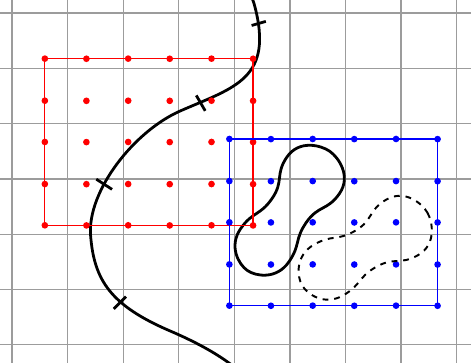}
  \mcaption{fig:grid_search}{}{A \twod depiction of the parallel candidate collision
  pair algorithm. Shown is the implicit spatial grid (gray), a piece of the
blood vessel $\Gamma$ (open black curve), an \rbc $\gamma_i$ at the current time
step (closed black curve) and at the next time step (dotted closed back curve).
Also shown is the space-time bounding box and bounding box samples of a single
patch (red square and red dots) and an \rbc (blue square and blue dots).
}
\end{minipage}
\end{figure}%
To solve the \lcp in \cref{step:lcp_solve}, we follow the approach
detailed in \cite[Section 3.2.2, Section 3.3]{lu2017}. 
We reformulate the problem first in standard \lcp form with diagonally-dominant system matrix $\vB$, then 
solve an equivalent root finding problem by applying a minimum-map Newton's
method. This can be restructured to use \gmres, so we only need to repeatedly apply  
$\vB$ to vectors to solve the \lcp.
Each entry $\vB_{ij}$ is the change in the $j$th contact volume induced by the
$k$th contact force, which is explicitly defined in
\cite[Algorithm 3]{lu2018parallel}.
This means that $\vB$ is of size $m \times m$, where $m$ is the number of
collisions, but is extremely sparse.
We need not store the entire matrix explicitly; we only compute the non-zero entries and
store them in a distributed hash-map.
Computing these matrix elements requires an accumulation of all coupled
collision contributions to the velocity, which requires just a sparse
\texttt{MPI\_All\_to\_Allv} to send each local contribution to the process containing
$V_i(t^{+,k})$.

An important step to ensure good scaling of our collision handling algorithm is
to minimize the number of triangle-vertex pairs that are found in
\cref{step:near_tri_pair}. One could explicitly compute an
all-to-all collision detection on all meshes in the system, but this requires
$O(N^2)$ work and global communication.
We perform a high-level filtering first to find local \textit{candidate collision
mesh pairs}, then only communicate and compute the required $O(m)$ information.
Since spatially-near mesh pairs may be on different 
processors, we need a parallel algorithm to compute these collision candidates.

To address this, we reuse \cref{step:bbnear,step:computeid,step:sort} from
\cref{sec:closest_point} and adapt it to this problem. For each mesh in the
system, we form the \textit{space-time
  bounding box} of the mesh: the smallest axis-aligned bounding box
containing the mesh at positions $\vX_i$ and $\vX_i^+$, as shown in
\cref{fig:grid_search}.  For patches $P_i$, note that $P_i^+ = P_i$.
This means one can reuse the bounding box of $P_i$ constructed in
\cref{sec:closest_point} for this purpose and simply set $d_\eps$ to zero.
After forming all space-time bounding boxes for the meshes of all patches and \rbcs, we
apply steps \cref{step:computeid,step:sort} directly to these boxes.
\cref{step:sort} will communicate meshes with the same spatial sorting key to
the same processor; these meshes are collision candidate pairs.
Once the computation is local and candidate collision pairs are identified, we
can proceed with the
\ncp solution algorithm described above.

\section{Results\label{sec:results}}

In this section, we present scalability results for our blood flow simulation
framework on various test geometries, simulations with various volume fractions 
and demonstrate the convergence behavior of our numerical methods.

\begin{figure}[!th]
\centering
\begin{minipage}[b]{0.48\textwidth}
\begin{tikzpicture}[scale=0.7] 
  \begin{semilogxaxis}[
    scale only axis,
    axis lines=left,
    xtick=data,
    xticklabels={384,768,1536,3072,6144,12288},
    ybar stacked,
    legend style={
      draw=none,
      at={(0.42,1.0)},
      anchor=north,
      legend columns=3,
      /tikz/every even column/.append style={column sep=0.5cm}},
    legend entries={\bf COL, \bf BIE-solve, \bf BIE-FMM, \bf
      Other-FMM, \bf Other},
    xlabel=\cpu cores $\rightarrow$,
    ytick={0,2000000,4000000,6000000,8000000,1000000},
    ymajorgrids,ylabel=wall-time $\times$ \cpu cores (bar) $\rightarrow$,
    xmin=250,
    xmax=25600,
    ymin=0,
    ymax=10000000,
    bar width=22pt,
    width=3.8in,
    height=3in]
 
    \addplot[color=black, fill=clr10]
    table[x=cs,y=col] {result_data/strong_result_scaled_detail};

    \addplot[color=black, fill=clr5]
    table[x=cs,y=wsolveother] {result_data/strong_result_scaled_detail};

    \addplot[color=black, fill=clr12]
    table[x=cs,y=wfmm] {result_data/strong_result_scaled_detail};

    \addplot[color=black, fill=clr13]
    table[x=cs,y=otherfmm] {result_data/strong_result_scaled_detail};

    \addplot[color=black, fill=clr14]
    table[x=cs,y=other] {result_data/strong_result_scaled_detail};

  \end{semilogxaxis}
\end{tikzpicture} 
\end{minipage}
\vfill
\begin{minipage}[b]{0.48\textwidth}
        \centering
        \begin{tabular}{ccccccc}\toprule
            cores  \!\!\!\!\!\!\!\!\!     &\!\!\! $384$ \!\!\! & $768$  & \!\!$1536$ & \!\!$3072$ &
            $6144$ & \!\!\! $12288$ \\ \cmidrule(lr){1-1} \cmidrule(lr){2-7}
            total time (sec)  & $11257$ & $5751$ & $3268$ & $1887$ & $1116$ & $718$ \\ 
            efficiency  & $1.00$  & $0.98$ & $0.86$ & $0.75$ & $0.63$ & $0.49$ \\ \hline
            {\bf COL}+{\bf BIE-solve} (sec)  & 3901  & 1843 & 1046 & 596  & 317  & 183 \\
            efficiency & 1.00  & 1.05 & 0.93 & 0.82 & 0.77 & 0.66 \\ \bottomrule
\end{tabular}
\end{minipage}
\mcaption{fig:sscale}{}{
  Strong scalability of a simultion with $40960$ \rbcs on Stampede's
  \abbrev{SKX} partition
  for the vessel network geometry shown in \cref{fig:strong-scale-domain}. The
  vessel is discretized with $40960$ polynomial patches. Shown in the bar graph
  is a breakdown of 
  the compute resources (wall-time $\times$ CPU cores) required by the
  individual components for a simulation with 10 time
  steps on 384 to 12288 cores. The compute resources used by the main algorithms presented in this
  paper are {\bf\em COL} (collision handling), {\bf\em BIE-solve}
  (computation of $\vu^{\Gamma}$, not including FMM calls). Shown in different gray scales are the
  compute resources required by FMM ({\bf\em BIE-FMM} and {\bf\em
    Other-FMM}) and other operations ({\bf\em Other}). Shown in the table
  are the compute time and the parallel efficiency for the
  overall computation and for the sum of {\bf\em COL} and {\bf\em
    BIE-solve}. For the collision avoidance and the
  boundary solve we observe a parallel efficiency of $66$\% for a
  32-fold increase from 384 to 12288 CPU cores.
}
\end{figure}
\subsection{Implementation and example setup}\label{ss:implementation}

\textbf{Architecture and software libraries. } We use the Stampede2 system at the Texas Advanced Computing Center (\abbrev{TACC})
to study the scalability of our algorithms and implementation.
Stampede2 has two types of compute nodes, the Knights Landing (\abbrev{KNL}) compute nodes
and the Skylake (\abbrev{SKX}) compute nodes.
The \abbrev{SKX} cluster has 1,736 dual-socket compute nodes, each with two
24-core 2.1\abbrev{GHz} \cpu{s} and 192\abbrev{GB} of memory.
The \abbrev{KNL} cluster has 4,200 compute nodes, with a 68-core Intel Xeon Phi
7250 1.4\abbrev{Ghz} \cpu{s} and 96\abbrev{GB} of memory plus 16\abbrev{GB} of
high-speed MCDRAM.
We run our simulations in a hybrid distributed-shared memory fashion: we run one
\abbrev{MPI} process per node, with one OpenMP thread per hardware core.
Our largest simulations use 256 \abbrev{SKX} and 512 \abbrev{KNL} nodes.

We leverage several high-performance libraries in our implementation.
We use \abbrev{PETSc}'s \cite{balay2017petsc} parallel matrix and vector operations, and its
parallel \gmres solver. 
Management and distribution of patches describing the blood vessel
geometry uses the \p4est library \cite{BursteddeWilcoxGhattas11}, and
we use \pvfmm
\cite{malhotra2015} for parallel \fmm evaluation.
We also heavily leverage \texttt{Intel MKL} for fast dense linear algebra
routines at the core of our algorithms and \texttt{paraview} for our
visualizations.

\begin{figure}[!htb]
\centering
\begin{minipage}[b]{0.48\textwidth}
\begin{tikzpicture}[scale=0.8] 
  \begin{semilogxaxis}[
    axis lines=left,
    xtick=data,
    xticklabels={48,192,768,3072,12288},
    ybar stacked,
    legend style={
      draw=none,
      at={(0.43,1.0)},
      anchor=north,
      legend columns=3,
      /tikz/every even column/.append style={column sep=0.5cm}},
    legend entries={\bf COL, \bf BIE-solve, \bf BIE-FMM, \bf
      Other-FMM, \bf other},
    xlabel=\cpu cores $\rightarrow$,
    ytick={0,5000,10000,15000},
    ymajorgrids,ylabel=wall-time $\rightarrow$,
    xmin=24,
    xmax=25600,
    ymin=0,
    ymax=15000,
    bar width=22pt,
    width=4.05in,
    height=3in]

    \addplot[color=black, fill=clr10]
    table[x=cs,y=col] {result_data/weak_result_large_grain_large_vol};

    \addplot[color=black, fill=clr5]
    table[x=cs,y=wsolveother] {result_data/weak_result_large_grain_large_vol};

    \addplot[color=black, fill=clr12]
    table[x=cs,y=wfmm] {result_data/weak_result_large_grain_large_vol};

    \addplot[color=black, fill=clr13]
    table[x=cs,y=otherfmm] {result_data/weak_result_large_grain_large_vol};

    \addplot[color=black, fill=clr14]
    table[x=cs,y=other] {result_data/weak_result_large_grain_large_vol};

  \end{semilogxaxis}
\end{tikzpicture} 
\end{minipage}
\vfill
\begin{minipage}[b]{0.48\textwidth}
        \centering
        \begin{tabular}{cccccc}\toprule
            cores           & $48$   & $192$  & $768$  & $3072$ & $12288$ \\ \cmidrule(lr){1-1} \cmidrule(lr){2-6}
            vol fraction    & $19\%$ & $20\%$ & $23\%$ & $26\%$ & $27\%$ \\ 
            \#collision/ \#\rbcs      & $15\%$ & $13\%$ & $17\%$ & $15\%$ & $16\%$ \\ \bottomrule
            total time (sec)  & $7070$ & $8892$ & $10032$ & $10869$ & $12446$  \\ 
            efficiency  & $ - $  & $1.00$ & $0.88$ & $0.81$ & $0.71$ \\ \bottomrule
            {\bf COL}+{\bf BIE-solve} (sec)   & 1461 & 2345 & 2926 & 3222 & 3904 \\
             efficiency & $-$ & 1.00 & 0.80 & 0.73 & 0.60 \\\bottomrule
        \end{tabular}
\end{minipage}
\mcaption{fig:wscale-large-grain}{}{Weak scalability on Stampede's
  \abbrev{SKX} partition with node grain size of $4096$ \rbcs and
  $8192$ polynomial patches per compute node (each node has 48 cores)
  for the vessel geometry shown in
  \cref{fig:wsscale-domain}. Increasing the number of \rbcs and
  boundary patches is realized by decreasing the size of the \rbcs as
  discussed in \cref{ss:weak}.  Shown in the bar graph is a breakdown of
  wall-time spent in individual components for a simulation with 10 time
  steps on 136 to 12288 cores (i.e., 4 to 256 nodes). The explanation
  of the labels used in the legend is detailed in
  \cref{fig:sscale}. Additionally, we show the volume fraction of
  \rbcs for each simulation, as well as the percentage of vesicles
  where the \rbc-\rbc or \rbc-vessel collision prevention is active.
  We report the parallel scalability with respect to 192 cores, as the
  smallest simulation is in a single node and no MPI communication is
  necessary. The largest simulation has 1,048,576 \rbcs and 2,097,152
  polynomial patches and an overall number of 
  3,042,967,552 unknowns per time step.
}
\end{figure}

\begin{figure}[!htb]
\centering
\begin{minipage}[b]{0.48\textwidth}
\begin{tikzpicture}[scale=0.8] 
  \begin{semilogxaxis}[
    axis lines=left,
    xtick=data,
    xticklabels={136, 544, 2176, 8704, 34816},
    ybar stacked,
    legend style={
      draw=none,
      at={(0.43,1.0)},
      anchor=north,
      legend columns=3,
      /tikz/every even column/.append style={column sep=0.5cm}},
    legend entries={\bf COL, \bf BIE-solve, \bf BIE-FMM, \bf
      Other-FMM, \bf other},
    xlabel=\cpu cores $\rightarrow$,
    ytick={0,1000,2000,3000,4000,5000,6000,7000},
    ymajorgrids,ylabel=wall-time $\rightarrow$,
    xmin=64,
    xmax=64000,
    ymin=0,
    ymax=7000,
    bar width=22pt,
    width=4.05in,
    height=3in]

   \addplot[color=black, fill=clr10]
    table[x=cs,y=col] {result_data/weak_knl_result_detail};

    \addplot[color=black, fill=clr5]
    table[x=cs,y=wsolveother] {result_data/weak_knl_result_detail};

    \addplot[color=black, fill=clr12]
    table[x=cs,y=wfmm] {result_data/weak_knl_result_detail};

    \addplot[color=black, fill=clr13]
    table[x=cs,y=otherfmm] {result_data/weak_knl_result_detail};

    \addplot[color=black, fill=clr14]
    table[x=cs,y=other] {result_data/weak_knl_result_detail};

  \end{semilogxaxis}
\end{tikzpicture} 
\end{minipage}
  \vfill
\begin{minipage}[b]{0.48\textwidth}
        \centering
        \begin{tabular}{cccccc}\toprule
            cores &                         $136$   & $544$  & $2176$  & $8704$ & $34816$ \\ \cmidrule(lr){1-1} \cmidrule(lr){2-6}
            vol fraction &               $17\%$ & $19\%$ & $20\%$ & $23\%$ & $26\%$ \\
            \#collision/ \#\rbcs&               $10\%$ & $15\%$ & $13\%$ & $17\%$ & $15\%$ \\\bottomrule
            total time (sec)  & $2739$ & $3203$ & $3768$ & $4782$ & $5806$  \\
            efficiency  & $1.00$  & $0.86$ & $0.73$ & $0.57$ & $0.47$ \\\bottomrule
            {\bf COL}+{\bf BIE-solve} (sec) &   $642$  & $808$ & $982$ & $1532$ & $1480$ \\
             efficiency &  $1.00$ & $0.79$ & $0.65$ & $0.42$ & $0.43$ \\\bottomrule
        \end{tabular}
\end{minipage}

\mcaption{fig:wscale-knl}{}{
  Same as \cref{fig:wscale-large-grain} but on Stampede2's
  \abbrev{KNL} partition with $512$ \rbcs and
  $1024$ vessel boundary patches per node (each node has 68
  cores). We find an overall parallel scalability of 47\% for a
  256-fold increase of the problem size.
}
\end{figure}

\begin{figure}[!htb]
\centering
\begin{minipage}[b]{0.2\textwidth}
    \includegraphics[angle=0,width=.98\linewidth]{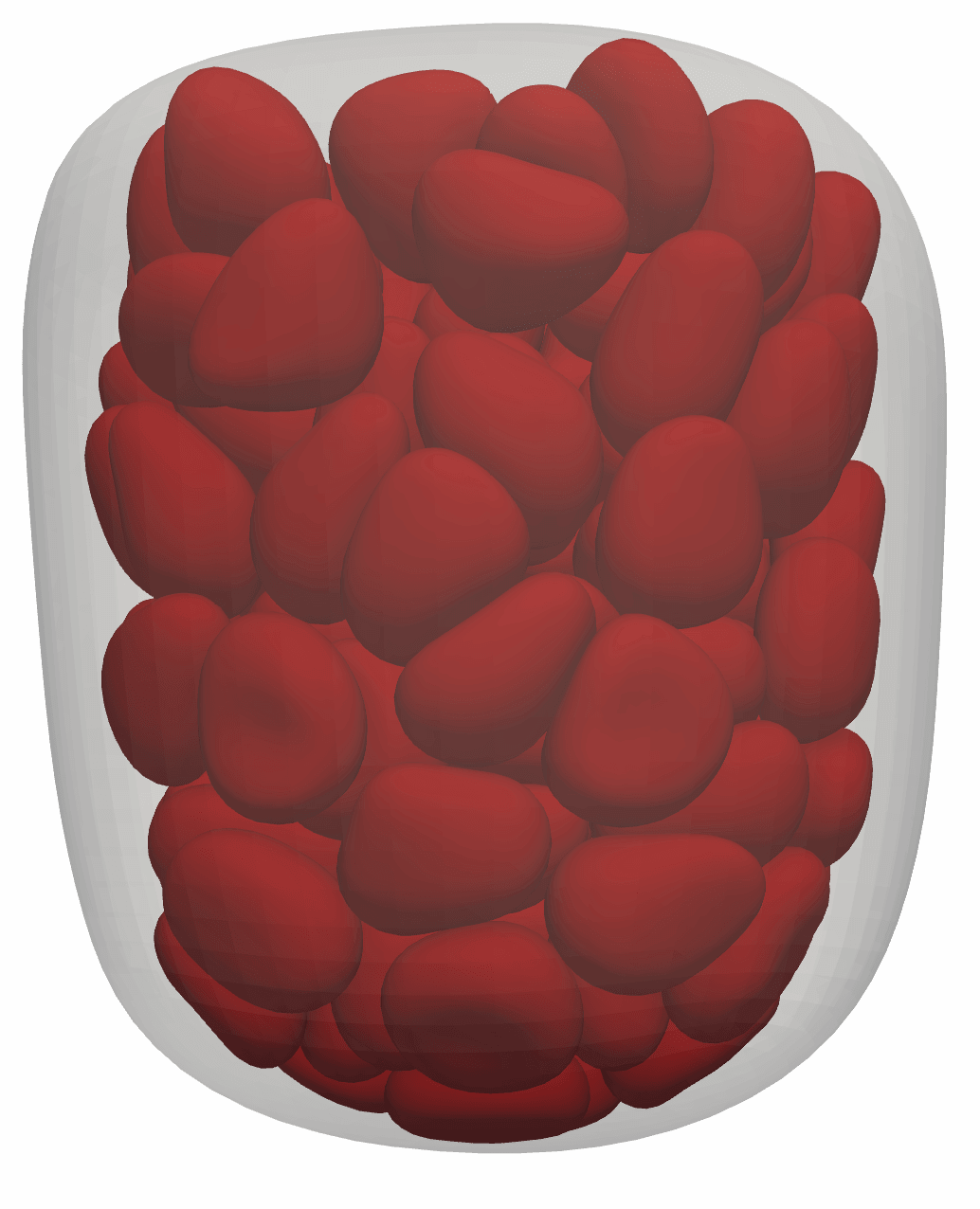}
\end{minipage}
\hspace{2ex}
\begin{minipage}[b]{0.2\textwidth}
    \includegraphics[angle=0,width=.98\linewidth]{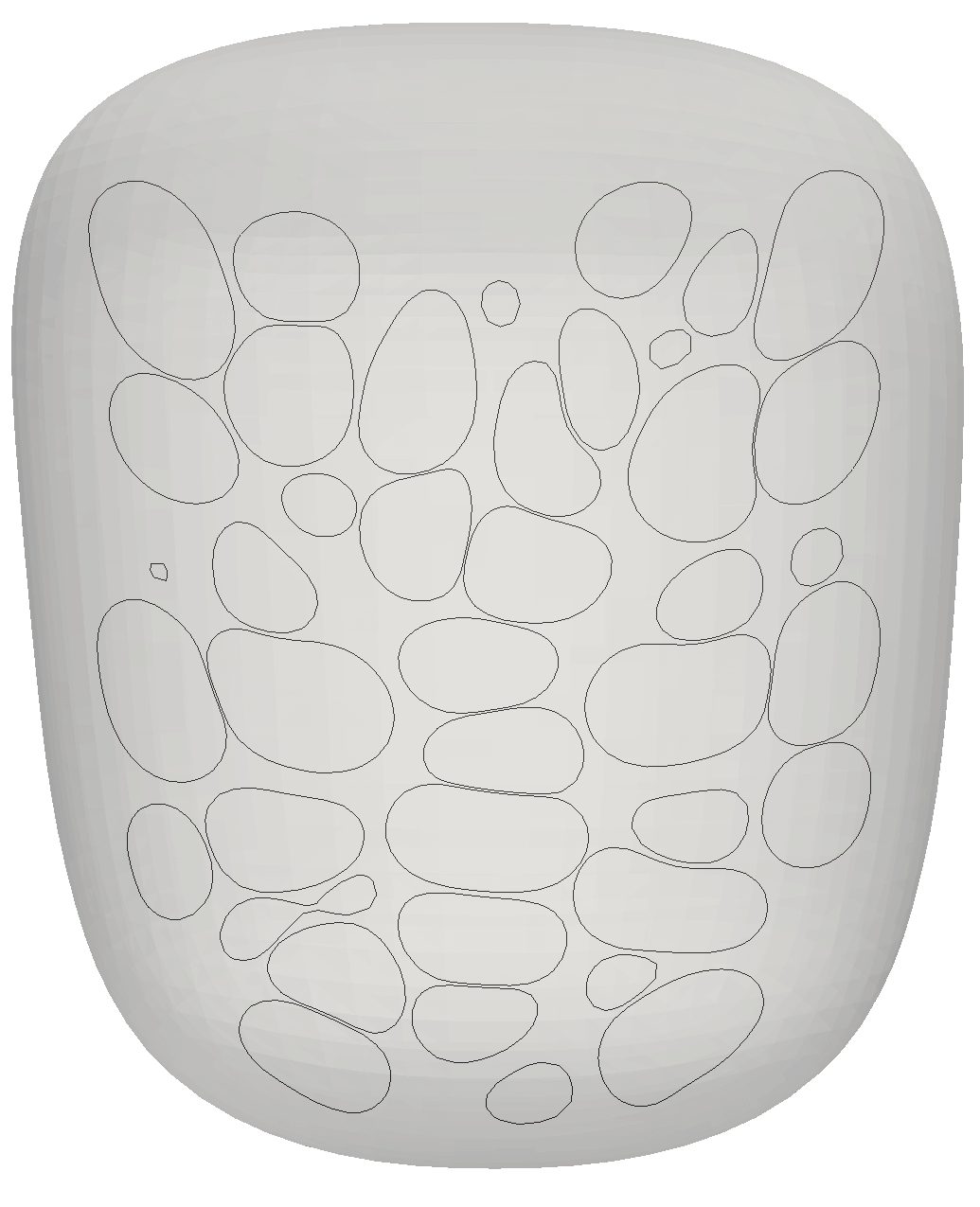}
\end{minipage}
\vfill
\begin{minipage}[b]{0.2\textwidth}
    \includegraphics[angle=0,width=.98\linewidth]{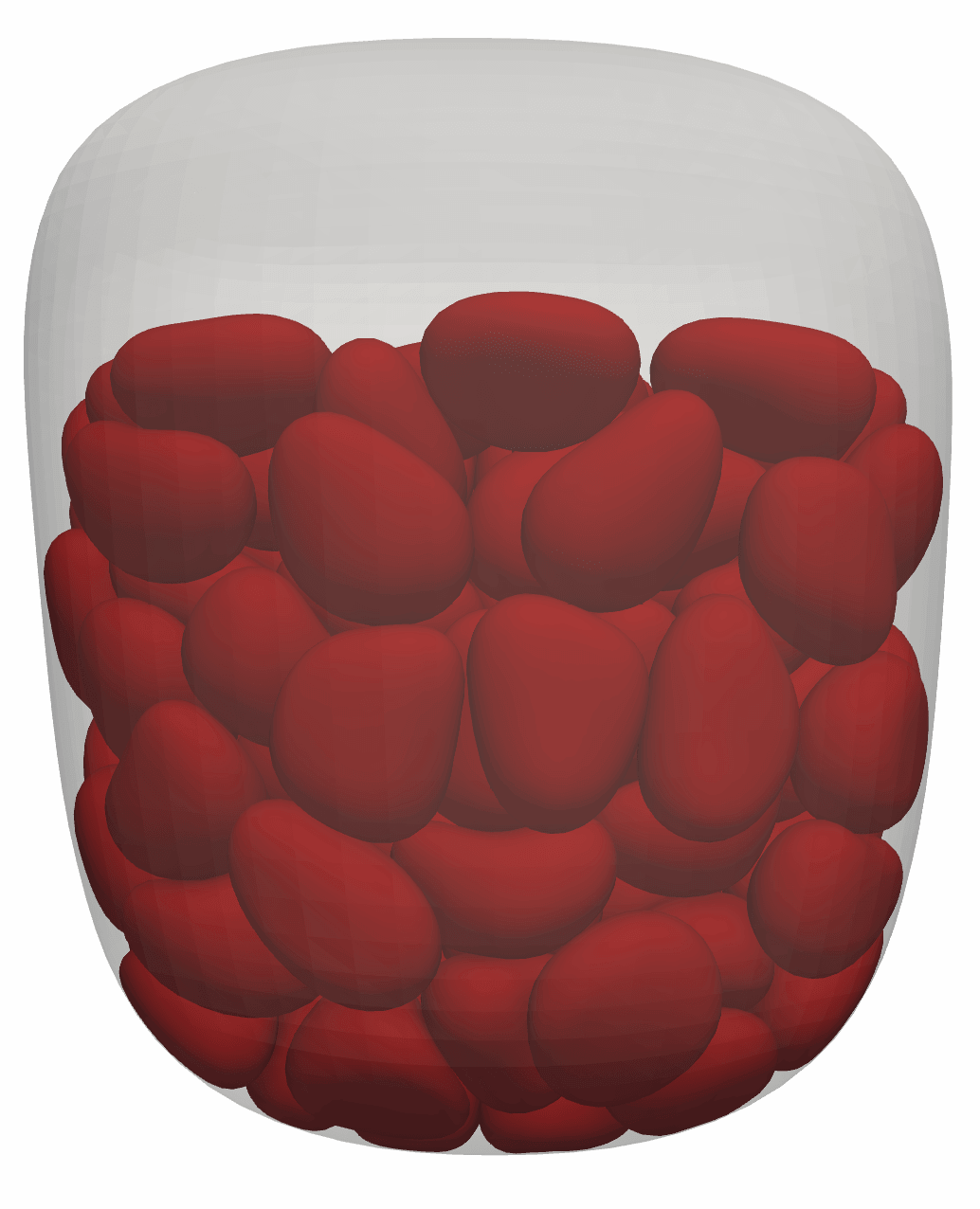}
\end{minipage}
\hspace{2ex}
\begin{minipage}[b]{0.2\textwidth}
    \includegraphics[angle=0,width=.98\linewidth]{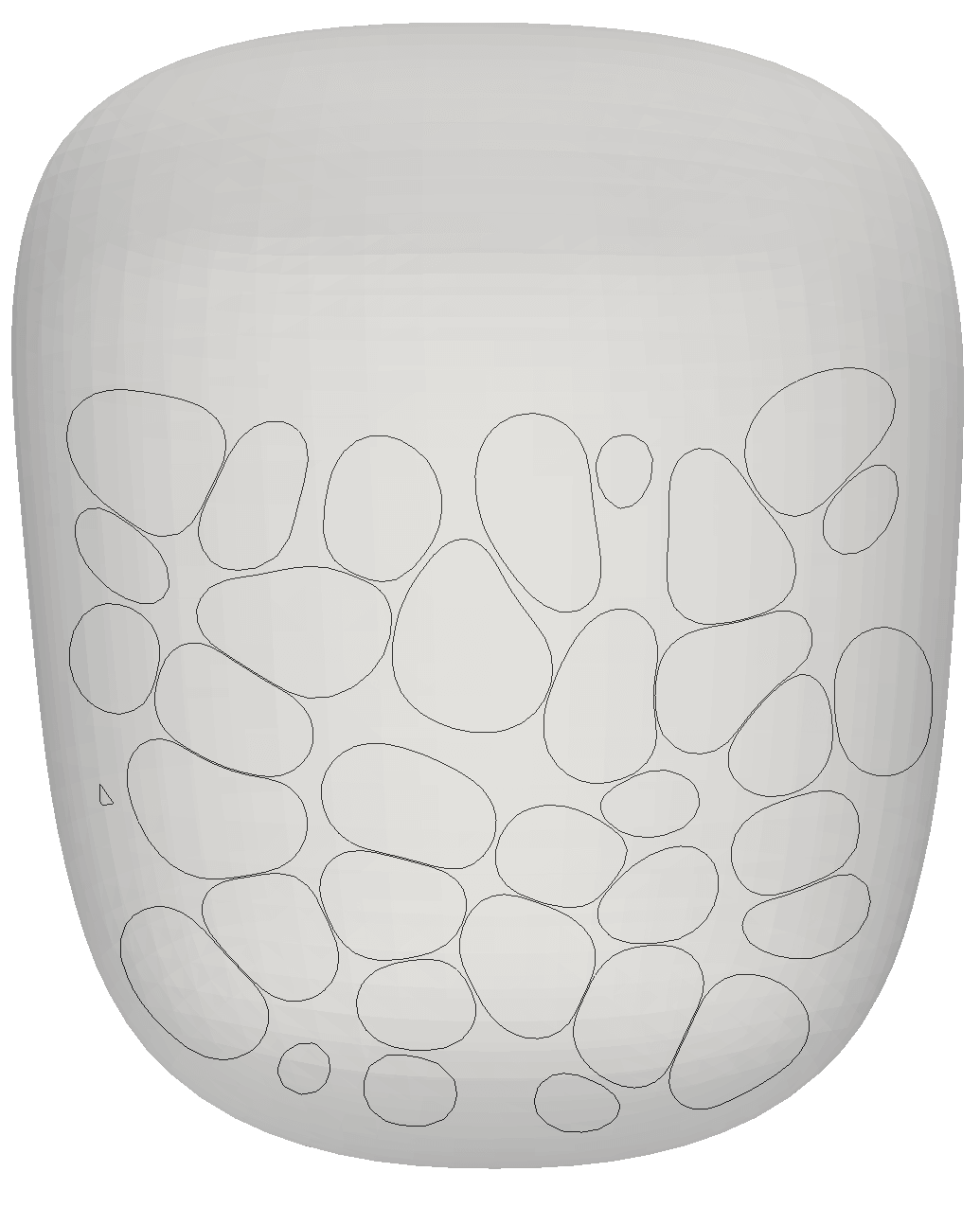}
\end{minipage}
    \mcaption{fig:high-vol-snap}{}{Shown is a high-volume fraction
      sedimentation due to gravitational force.
    The initial configuration (top figures) has a volume fraction of $47\%$. 
    As the cells sediment to the lower part of the domain (bottom figures), the local
    volume fraction of the final state in this lower part of the
    domain is around $55\%$. Shown on the
    right side are slices through the center of the domain together
    with the \rbc boundaries in the initial and final configuration.
    The full simulation video is available at
    \href{https://vimeo.com/329509435}{\textcolor{blue}{https://vimeo.com/329509435}}.
}
\end{figure}
\textbf{Discretization and example setup. }
For all test cases we present, we discretize each \rbc with $544$ quadrature
points and 2,112 points for collision detection.
The blood vessel geometry is represented with 8th order tensor-product
polynomial patches with  121 quadrature points per patch and 484
equispaced points for collision detection.
The parameters chosen for singular/near-singular integration are $p=8$ and $\eta=1$, with $R=r=.15L$ for strong scaling tests and $R=r=.1L$ for weak scaling tests.
The value of $L$ is the square root of the surface area of the patch containing the closest point to the target, called the \emph{patch size}; this choice allows for a consistent extrapolation error over the entirety of $\Gamma$.

Since our scaling tests are performed on complex, realistic blood
vessel geometries, we must
algorithmically generate our initial simulation configuration.
We prescribe portions of the blood vessel as inflow and outflow regions and
appropriately prescribe positive and negative parabolic flows (inlet and outlet
flow) as boundary
conditions, such that the total fluid flux is zero.
To populate the blood vessel with \rbcs, we uniformly sample the volume of the 
bounding box of the vessel with a spacing $h$ to find point locations
inside the domain at which we place \rbcs in a random orientation.
We then slowly increase the size of each \rbc until it collides with the
vessel boundary or another \rbc; this determines a single \rbc's size.
We continue this process until all \rbcs stop expanding; this means that 
we are running a simulation of \rbcs of various sizes.
We refer to this process as \textit{filling the blood vessel with }\rbcs.
This typically produces \rbcs of radius $r$ with $r_0 < r < 2r_0$ with $r_0$ chosen proportional to $h$. 
This is a precomputation for our simulation, so we do not include this
step in the timings we report for weak and strong scaling.
We emphasize that these simulations are primarily for scaling purposes
of our algorithms and are not expected to represent true blood flows.
The platform can of course be applied to length scales where viscous
flow is a valid assumption.

Additionally, \rbcs in such a confined flow will collide with the
blood vessel wall
if special care is  not taken near the outflow part of the boundary.
We define regions near the inlet and outlet flows where we can safely add and
remove \rbcs.
When an \rbc $\gamma_i$ is within the outlet region, we subtract off the velocity due to
$\gamma_i$ from the entire system and move $\gamma_i$ into an inlet
region such that the arising \rbc
configuration is collision-free.

\textbf{Limiting \abbrev{GMRES} iterations. }
We have observed that the GMRES solver typically requires 30 iterations
or less for convergence for almost all time steps, but the number
of needed iterations may vary more in the first steps. 
To simulate the amount of work in a typical simulation time
step, we cap the number of \gmres iterations at 30 and report weak and strong
scaling for these iterations. A more detailed analysis of this behavior is needed.


\begin{figure*}[!bht]
\centering
  \includegraphics[angle=0,width=.85\linewidth]{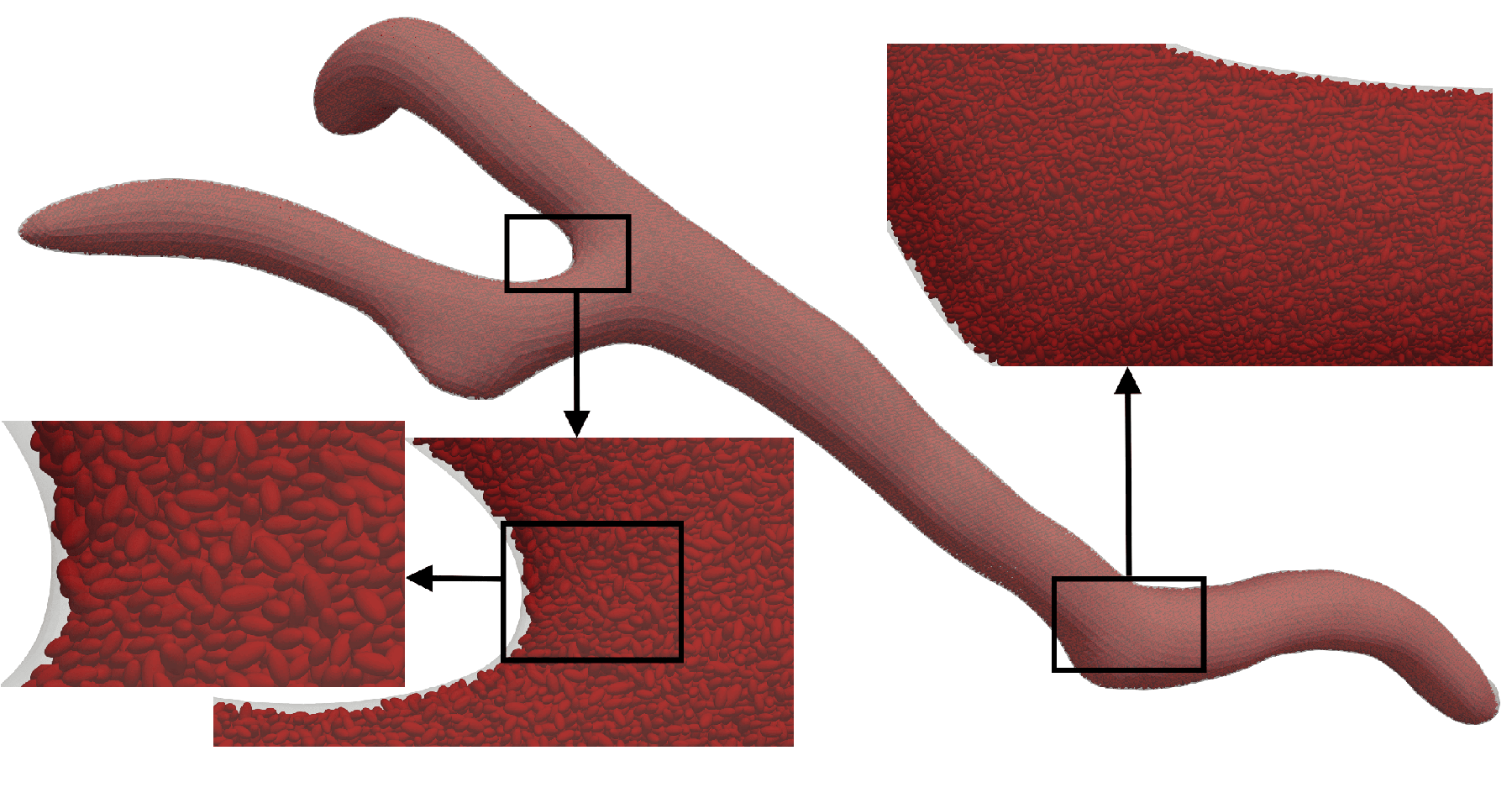}
  \mcaption{fig:wsscale-domain}{}{
    For our weak scaling experiments, we use the the vessel geometry
    shown above with inflow boundary conditions on the
    right side and outflow boundary condition on the two left sides.
    To setup the problem, we fill the vessel with 
    nearly-touching \rbcs of different sizes to obtain a desired
    number, and refine the
    vessel geometry patches. The figure above shows a setup with overall
    262,144 \rbcs at a volume fraction of $26\%$. 
  }
\end{figure*}

\subsection{Parallel scalability}\label{ss:scalability}

  Here, we present strong and weak scalability results for our \rbc simulations. 
  We decompose the time required for a complete simulation into the following categories:
 \begin{itemize}
    \item {\bf\em COL}: detection and resolution of collisions among
      \rbcs and between \rbcs and the vessel walls;
      \item {\bf\em BIE-solve}: computing $\vu^{\Gamma}$, not including \fmm calls. This includes all of the steps for singular/near-singular integration in \cref{sec:solver} except the evaluation $\vu^\Gamma$ at the check points.
      \item {\bf\em BIE-FMM}: \fmm calls required to evaluate $\vu^{\Gamma}$ at the check points and at points on \rbcs
      \item {\bf\em Other-FMM}: \fmm calls required by other algorithms 
      \item {\bf\em Other}: all other operations
  \end{itemize}
In the discussion below, we focus on {\bf\em COL} and {\bf\em BIE-solve}, as
  they are the primary algorithmic contribution of this work, and discuss
  how to reduce the computational time required for {\bf\em BIE-FMM}.

\textbf{Strong scalability. }\label{ss:strong}
To study the strong scalability of our algorithms, we use the blood vessel
geometry and \rbc configuration in
\cref{fig:strong-scale-domain}-left.
This simulation contains 40,960 \rbcs and the blood vessel is
represented with 40,960 patches.
With four degrees of freedom per \rbc quadrature point and three per vessel
quadrature point, this amounts to
89,128,960 and 14,868,480 degrees for the \rbcs and blood vessel, respectively 
(103,997,440 in total).
As can be seen from \cref{fig:sscale}, we achieve a $15.7$-fold
speed-up in total wall-time scaling from 384 to 12288 cores,
corresponding to $49\%$ parallel efficiency.  This level of  parallel
efficiency is partially due to the calls to the fmm library
\pvfmm. The strong scalability of \pvfmm we observe is largely
consistent with the results reported in \cite{malhotra2016algorithm}.
Neglecting the time for calls to \fmm, i.e., only counting the time
for the boundary solver to compute $\vu^\Gamma$ and for collision
prevention, we find $66$\%
parallel efficiency when scaling strongly from 384 to 12288 cores.
We see that the parallel collision handling and integral equation solver
computations, excluding \fmm, scale well as the number of cores is increased.

\textbf{Weak Scalability. }\label{ss:weak}
Our weak scalability results are shown in
\cref{fig:wscale-large-grain,fig:wscale-knl}.
Both tests are performed on the blood vessel displayed in
\cref{fig:wsscale-domain}.
We use an initial boundary composed of a fixed number $M$ of polynomial
patches and fill the domain with roughly $M/2$ \rbcs (which requires spacing
$h$).
To scale up our simulation by a factor of four, we: (1) subdivide the $M$ polynomial
patches into $4M$ new but equivalent polynomial patches (via subdivision rules
for Bezier curves); (2) refill the domain with \rbcs using spacing 
$h/\sqrt[3]{4}$. This places $2M$ \rbcs in the domain volume.
We repeat this process each time we increase the number of cores by a factor of
four in order to keep the number of patches and \rbcs per core constant.
In the tables in  \cref{fig:wscale-large-grain,fig:wscale-knl},
we report parallel efficiency with respect to the
first multi-node run on both \skx and \knl architectures, i.e, with
respect to 192 and 136 cores, respectively.

The largest weak scaling test contains 1,048,576 \rbcs and 2,097,152 polynomial
patches on the blood vessel; we solve for 3,042,967,552 unknowns at each time
step and are able to maintain a collision-free state between
4,194,304,000 triangular surface elements at each time step.
Comparing the weak scalability results for \abbrev{SKX}
(\cref{fig:wscale-large-grain}) and \abbrev{KNL}
(\cref{fig:wscale-knl}), we observe similar qualitative
behavior. Note that the smallest test on the \abbrev{SKX} architecture
only uses a single node, i.e., no MPI communication was needed. This
explains the increased time for the collision prevention algorithms when going
from 1 (48 cores) to 4 nodes (192 cores). 
Note also that the simulation on the \abbrev{KNL} architecture
used a significantly lower number of \rbcs and geometry patches per
node. Thus, this simulation has a larger ratio of communication to
local work. This explains the less perfect scalability compared to
the results obtained on the \abbrev{SKX} architecture.
As with strong scaling, we see good parallel scaling of the non-\fmm-related parts
of the computation of $\vu^\Gamma$ and the collision handling algorithm.

Note that there is a slight variation in the number of collisions for
the run on 8704 cores on \knl.
This is an artifact of the \rbc filling algorithm.
Since we place \rbcs in random orientations and distribute \rbcs
randomly among processors, we do not have complete control over the percentage
of collisions or the volume fraction for each simulation in
\cref{fig:wscale-large-grain,fig:wscale-knl}, as can be seen from
the tables under these figures.
This can affect the overall scaling: For the run on 8704 cores, the
percentage of collisions is larger, explaining the longer
time spent in {\bf\em COL}.
Despite this phenomenon, we achieve good weak scaling overall.

\textbf{Discussion. }
The parts of the algorithm introduced in this paper scale as well as
or better than the \fmm implementation we are using.  However, our
overall run time is diminished by the multiple expensive
\fmm evaluations required for solving \cref{eq:double_layer_int_eq}.
This can be addressed by using a \textit{local} singular quadrature scheme,
i.e., compute a singular integral using the \fmm on
\cref{eq:smooth_double_layer_int_eq_patches_disc} directly, then compute a
singular correction locally.
This calculation has a three-fold impact on parallel scalability: (1) the \fmm
evaluation required is proportional to the size of the coarse discretization
rather than the fine discretization ($O((p+1)N)$ vs.\ $O((k+p)N)$);
(2) after the \fmm evaluation, the local correction is embarrassingly parallel;
(3) the linear operator \cref{eq:sing-quad} can be precomputed, making the
entire calculation extremely fast with \abbrev{MKL} linear algebra routines.
These improvements together will allow our algorithm to scale well beyond the
computational regime explored in this work.

\subsection[]{Verification}
There are few analytic results known about \rbcs in confined Stokes
flows against which we can verify our simulations. However,
exact solutions can be obtained for a part of our setup, invariants
(e.g., surface area) can be considered and solutions for smaller examples can
be verified against solutions with fine spatial and temporal discretizations.
In particular, in this section, we demonstrate the accuracy of the parallel boundary
solver presented in \cref{sec:solver} and numerically study the collision-free
time-stepping in \cref{sec:parallel-contact}. 

\textbf{Boundary solver. }
The error of the boundary integral solver is determined
by the error of integration and the GMRES error, the latter not depending on the number of discretization points due to good conditioning of the equation.  The integration error, in turn, can be separated into smooth quadrature error and interpolation error. The former is  high-order accurate \cite{trefethen2013approximation}.
Although our extrapolation is ill-conditioned, we observe good accuracy for $p\leq 8$. The singular evaluation in \cref{sec:solver} converges with rate
$O(L^p + L^q)$ corresponding to $p$th order extrapolation and $q$th
order quadrature.
To confirm this numerically, we solve an interior Stokes problem on the
surface in \cref{fig:boundary-err}-right.
We evaluate a prescribed analytic solution at the discretization
points to obtain the boundary condition.
We then solve \cref{eq:int_eq_disc} and compare the numerical
solution at on-surface samples different from discretization points, evaluated using the algorithms of \cref{sec:solver}. 
We use $\eta =2$, $q=16$, $p=8$, $R=.04\sqrt{L}$ and $r = R/8$.
In \cref{fig:boundary-err}-left, we report the relative error in the
infinity-norm of the velocity.
By choosing check point distances proportional to $\sqrt{L}$,
we observe the expected $O(L^7)$ convergence.
\begin{figure}[h]
  \centering
\begin{minipage}[b]{0.3\textwidth}
\begin{tikzpicture}
  \begin{loglogaxis}[%
    width=1.75in,
    height=1.25in,
      scale only axis,
      separate axis lines,
      every outer x axis line/.append style={white!15!black},
      every x tick label/.append style={font=\color{white!15!black}},
      xlabel={Max patch size},
      every outer y axis line/.append style={white!15!black},
      every y tick label/.append style={font=\color{white!15!black}},
      ylabel={Max relative error},
      ytick={1e-1,1e-3,1e-5, 1e-7, 1e-9},
      xlabel style={font=\small},
      ylabel style={font=\small},
      legend pos=north west,
      legend cell align=left,
    ]

    \addplot [color=plt-orange,line width=1.0pt,mark=square*,mark size=1.2pt,mark repeat=1,mark phase=0]
        table[x index=0,y index=1]{figs/bie-conv.dat};
        \addlegendentry{Max rel. error}

    \addplot [color=black,line width=1.0pt,dotted,domain=5e-2:.4]{400*x^7};
    \addlegendentry{$O(L^7)$}
  \end{loglogaxis}
\end{tikzpicture}%
\end{minipage}%
\hspace*{.1\linewidth}
\begin{minipage}[b]{0.4\textwidth}
\includegraphics[angle=0,width=.3\linewidth]{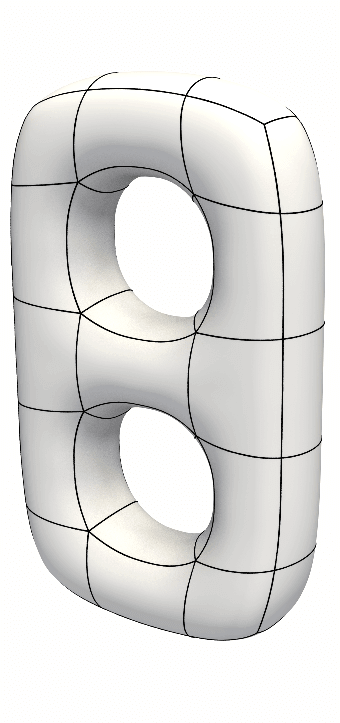}
\end{minipage}%
\mcaption{fig:boundary-err}{}{
  Error convergence test solving \cref{eq:int_eq_disc}. 
  We evaluate a known solution on the coarse discretization and solve for $\phi$.
On the left, we plot the maximum relative error in the infinity norm
of $\vu^\Gamma$ evaluated on the surface. 
On the right, we show the coarse discretization of the domain boundary
and patches.
}  \vspace{-5pt}
\end{figure}

\textbf{\rbcs with collision resolution and convergence. }
Our choice of \rbc  representation and discretization is spectrally
accurate in space for the approximation, differentiation and
integration of functions on \rbc  surfaces, as shown in
\cite{Veerapaneni2011}. Although we use first-order time-stepping in
this work,   \emph{spectral deferred correction} (\abbrev{SDC}) can be
incorporated into the algorithm exactly as in the 2D version described
in  \cite{lu2017}. This present work demonstrates second-order convergence in
time; however, \abbrev{SDC} can be made arbitrarily high-order
accurate.

For collision-resolution accuracy verification, we study the
convergence of our contact-free time-stepping with two \rbcs in shear
flow. As shown in \cref{fig:shear-snap}, at $T=0$, two \rbcs are
placed in a
shear flow $\vu = [z,0,0]$ in free-space. 
We first compute a reference solution without
collision handling  but with expensive adaptive fully implicit time-stepping to
ensure accurate resolution of the lubrication
layer between \rbcs. This reference simulatation used spherical
harmonics of order 32 and the time step had to be reduced to
6.5e-4 to prevent collisions.
In \cref{fig:shear-conv}, we show the convergence for the error in the centers of mass of each \rbc as a function of the time-step size. 
We use spherical harmonic orders $16$ and $32$ for the spatial
discretization to demonstrate the dominance of the time-stepping error. 
We observe first-order convergence with our locally-implicit backward Euler scheme which confirms that our collision resolution algorithm does not have a significant impact on time-stepping accuracy.
\begin{figure}[!htb]
  \subfloat[$t=0$\label{sfg:shear-snap1}]{\framebox(52,35){\includegraphics[width=52pt,height=35pt]{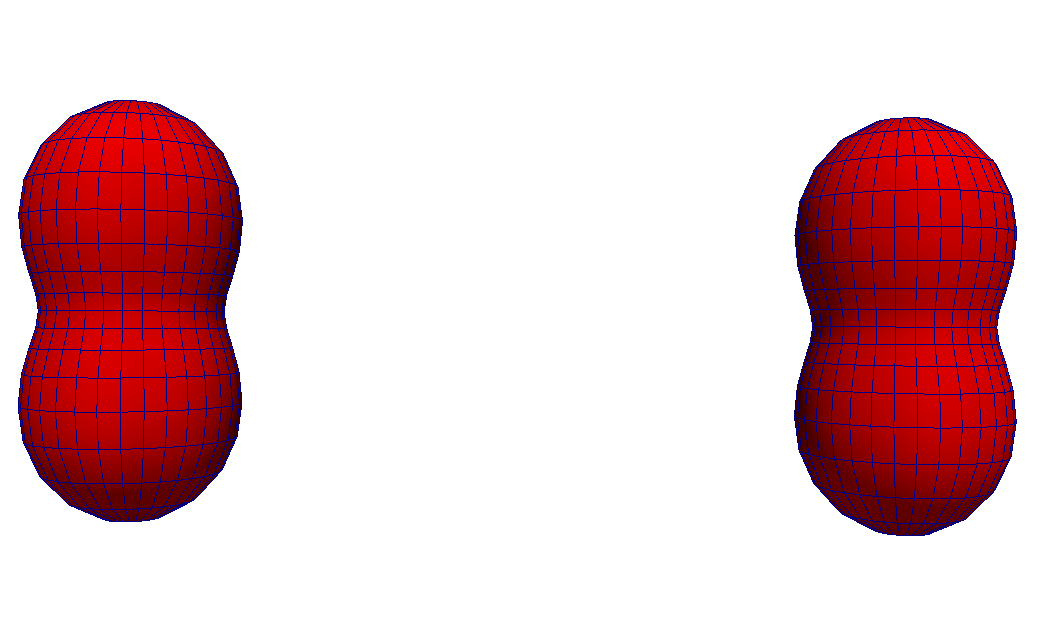}}}
  \hspace{1pt}
  \subfloat[$t=5$\label{sfg:shear-snap2}]{\framebox(52,35){\includegraphics[width=52pt,height=35pt]{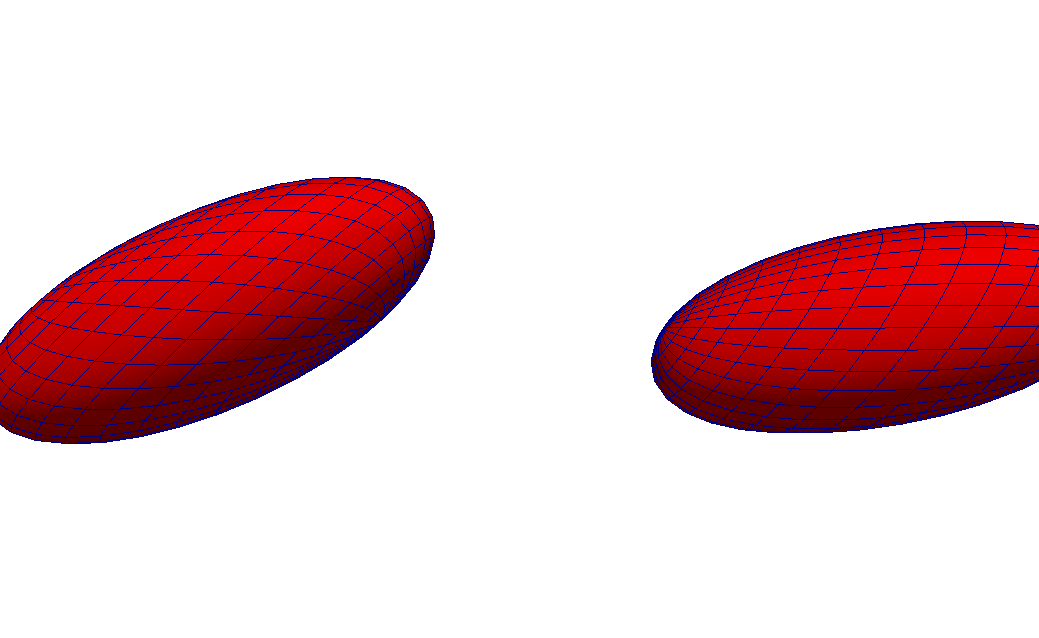}}}
  \hspace{1pt}
  \subfloat[$t=20$\label{sfg:shear-snap3}]{\framebox(52,35){\includegraphics[width=52pt,height=35pt]{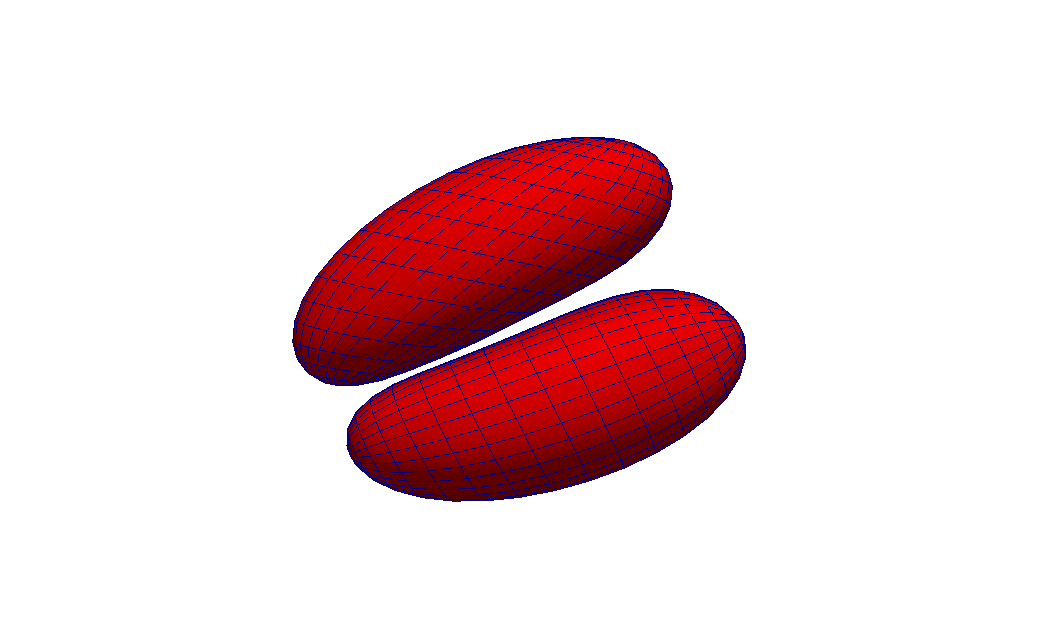}}}
  \hspace{1pt}
  \subfloat[$t=25$\label{sfg:shear-snap4}]{\framebox(52,35){\includegraphics[width=52pt,height=35pt]{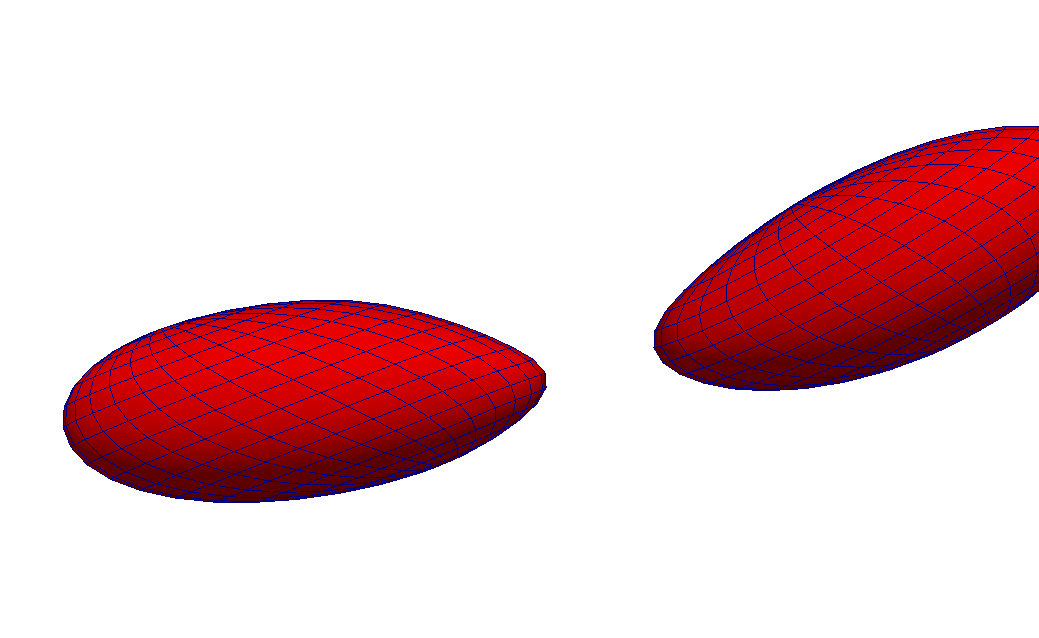}}}
  \mcaption{fig:shear-snap}{}{Snapshots of two vesicles in shear flow.} \vspace{-5pt}
\end{figure}%
\begin{figure}[!htb]
  \centering
    \includepgf{conv}
  \mcaption{fig:shear-conv}{}{
    Shown is the error in the final ($T=25$)
    centroid location as we decrease the time step size for two
    spherical harmonic orders $16$ and $32$. 
  We observe $O(\Delta t)$ convergence in time and hence the
  collission detection algorithm converges at the same order as the
  time stepper.}
  \vspace{-5pt}
\end{figure}%
\subsection{High volume fraction}
The \rbc volume fraction, i.e., the ratio of volume occupied by \rbcs
compared to the overall blood volume is 36-48\% in healthy women and
40-54\% in healthy men \cite{billett1990hemoglobin}. As can be seen in the tables in
\cref{fig:wscale-large-grain,fig:wscale-knl}, the volume fraction in
our weak scaling simulations is below these values, which is mostly due to
the procedure used to fill the blood vessel with \rbcs (see the
discussion in \cref{ss:implementation}).
However, \rbc volume fractions in capillaries and small arteries is
known to be be around 10-20\%
\cite{wang2013simulation,saadat2019simulation}, which our scaling
simulations achieve.
To demonstrate that we can simulate even higher volume fraction blood flows,
\cref{fig:high-vol-snap} shows a test of 140 \rbcs sedimenting under a
gravitational force in a small capsule. The volume fraction for this example is
47\%, calculated by dividing the amount of volume occupied by \rbcs by the
volume of the capsule.
By the end of the simulation, we achieve a volume fraction
of 55\% in the lower part of the domain (determined by bounding the
\rbcs by a tighter cylinder than the original domain boundary) since
the \rbcs have become more tightly packed.
While such high volume fractions typically do not occur in capillary flow on average, in some scenarios (local fluctuations, sedimentation, microfluidics) these high concentrations need to be handled.

\section{Conclusion\label{sec:conclusion}}
We have shown that our parallel platform for the simulation of red
blood cell flows is capable of accurately resolved long-time
simulation of red blood cell flows in complex vessel networks. We are
able to achieve realistic cell volume fractions of over 47\%, while avoiding
collisions between cells or with the blood vessel
walls. Incorporating blood vessels into red blood cell
simulations requires solving a
boundary integral equation, for which we use \gmres.
Each \gmres iteration computes a matrix-vector product,
which in turn involves singular quadrature and an \fmm
evaluation; the latter dominates the computation time. To avoid collisions,
we solve a nonlinear complementarity problem in the implicit part of
each time step. This requires repeated assembly of sparse matrices that, in
principle, couple all cells globally. Nevertheless, solving this
complementarity system yields close-to-optimal strong
and weak scaling in our tests. Overall, the vast majority of compute
time is spent in \fmm evaluations, which implies that the scaling behavior of our
simulation is dominated by the scalability of the \fmm implementation.
As discussed at the end of \cref{ss:scalability}, in the
future, we will employ a local singular quadrature scheme that will
allow us to significantly reduce the time spent in \fmm evaluations. This
will not only speed up the overall simulation but
also improve the weak and strong scalability of our simulation platform.

\section{Acknowledgements}
We would like to thank 
Dhairya Malhotra, 
Michael Shelley and 
Shenglong Wang
for support and various discussions throughout about various aspects of this work.
This work was supported by the US National Science Foundation
(\abbrev{NSF}) through grants DMS-1821334, DMS-1821305, DMS-1320621, DMS-1436591 and EAR-1646337.
Computing time on TACC's Stampede2 supercomputer was provided through the Extreme Science and
Engineering Discovery Environment (XSEDE), which is supported by
National Science Foundation grant number ACI-1548562.

\clearpage
\printbibliography

\end{document}